\documentclass[aps,twocolumn]{revtex4}
\usepackage{graphicx}
\usepackage{bm}

\begin{document} 

\title{\bf Non-equilibrium phase transitions in biomolecular signal
  transduction}

\author{Eric Smith$^{(1)}$, Supriya Krishnamurthy$^{(1,2,3)}$, Walter
  Fontana $^{(1,4)}$ and David Krakauer$^{(1)}$}

\affiliation{(1):Santa Fe Institute, 1399 Hyde Park Road, Santa Fe, NM
87501, USA \\ (2): Department of Physics, Stockholm University, SE- 106 91, Stockholm, Sweden \\ (3): School of Computer Science and Communication, KTH, SE- 100 44 Stockholm, Sweden \\ (4): Department of Systems Biology, Harvard Medical School, 200 Longwood Avenue, Bostom MA 02115, USA}

\date{\today}
\begin{abstract}

We study a mechanism for reliable switching in biomolecular
signal-transduction cascades.  Steady bistable states are created by
system-size cooperative effects in populations of proteins, in spite
of the fact that the phosphorylation-state transitions of any
molecule, by means of which the switch is implemented, are highly
stochastic.  The emergence of switching is a nonequilibrium phase
transition in an energetically driven, dissipative system described by
a master equation.  We use operator and functional integral methods
from reaction-diffusion theory to solve for the phase structure, noise
spectrum, and escape trajectories and first-passage times of a class
of minimal models of switches, showing how all critical properties for
switch behavior can be computed within a unified framework.

\end{abstract}

\maketitle

\section{Introduction}

\subsection{The emergence of devices from biomolecular systems}

Two related questions define a fundamental role for statistical
physics in systems biology: (1).~How do biomolecular systems achieve
reliable ``device-level'' behavior when they consist of highly
stochastic componentry~\cite{vonNeumann:problog:56}, in particular
molecular complexes held together by low-energy hydrogen or Van der
Waals bonds?  (2).~ Are such systems structured in a modular fashion
~\cite{vonDassow:SPN:00}, {\it i.e.}
can complex molecular networks decompose into
quasi-autonomous functional units performing identifiable tasks
which are robust against many physical parameter changes and
are recombinable in evolution?

The device logic of biomolecular systems employs many of the same
abstractions as electronic engineering~\cite{Sauro:proteomics:04},
including amplification, filtering, and switching.  Switches (which
use elements of amplification and filtering) are used wherever a
discrete sequence of events is required, enabling committed threshold
response to environmental cues during
development~\cite{Ferrell:xenopus:99,Ferrell:switch:99}, establishing
``checkpoints'' for intermediate states, in processes like cell
division that require strict sequencing of
events~\cite{Tyson:CellCycle:02}, or programming the complex
progressions of cell-type differentiation~\cite{vonDassow:SPN:00}.

Two classes of biomolecular processes in which modular switching is
widely recognized are gene expression and signal transduction.  In
gene expression, the switch states are associated with patterns of
genes activated for protein production, and a stochastic component is 
introduced in the system by the small number of weakly-bound
transcription factors~\cite{Sasai:gene_exp:03}.  In
signal transduction, the states of the switch are frequently
determined by the number of phosphate or methyl groups covalently
attached to specific target amino acid residues on one or more
dedicated proteins~\cite{Goldbeter:hypersensitivity:81}. Stochasticity
in this system can again be caused by the small number of proteins or
the complexes these proteins form  with the catalysts which promote 
attachment or detachment of the phosphate or methyl groups.  
Gene expression changes cell type on
slower timescales (minutes to hours) than signal transduction
(seconds), but since many cell-type changes occur in response to
external signals~\cite{Ferrell:xenopus:99,Ferrell:switch:99}, or rely on
amplification and stabilization of internal signals~\cite{Tyson:CellCycle:02},
switching behavior is often produced by both systems acting together.

In this paper we consider the problems of mechanism for the emergence
of robust switching in stochastic molecular systems, and of
quantitative estimation of the noise and stability properties of
switches.  We consider switching in signal transduction via the mechanism of 
phosphorylation  (addition of phosphate groups via the action of a kinase) 
and dephosphorylation (removal of phosphate groups by the action of a 
phosphatase) since these
are a common motif found in most if not all signalling networks.
In addition, the relative simplicity  
of phosphorylation transitions lends itself to the
abstraction of many real transduction cascades in terms of a few
processes, allowing us to isolate the problems of formation, control,
and robustness of the switch.  We observe that the most fundamental
unit in signal-transduction cascades is a single type of target
protein with multiple phosphorylation states (called phosphoepitopes),
among which transitions are naturally modeled as a reaction-diffusion
process.  The network properties necessary for switching, which may be
distributed in real systems among several proteins in a
signal-transduction cascade, or between the cascade and the genetic
transcription factors for the target proteins, are readily lumped together 
and assigned to
a single species to produce minimal models.  This coarse-graining reduces
network complexity while still keeping its essential regulatory features.
We are able to write
down exact master equations for such ideal models, and to solve them
systematically with field-theoretic methods from reaction-diffusion
theory.

\subsection{Senses of ``switching'', their uses, and how they are
  achieved robustly} 

Switching in biomolecular systems, at the least, refers to sigmoidal
response to input signals, termed
``ultrasensitivity''~\cite{Huang:MAPK:96}.  Such response implies a
sharp sigmoidal but continuous response in the concentration of a
molecule over a narrow range of a (stationary) signal.  In contrast,
{\it bistability\/} \cite{Tyson:Sniffers:03} is a form of switching
made possible when two stable states, $S_1$ and $S_2$, co-exist over a
signal range.  As a consequence, bistable systems exhibit two distinct
thresholds as the signal is varied, one at which a transition occurs
from $S_1$ to $S_2$ and another at which the system switches back from
$S_2$ to $S_1$. The separation of thresholds leads to path dependence
or hysteresis, and makes a switched state more impervious to
stochastic fluctuations of the signal around the transition point,
than in the ultrasensitive case. 
Hysteresis may be obtained by
adding positive feedback to sigmoidal
response~\cite{Ferrell:bistability:01}.  Weak hysteresis may be used
to stabilize input signals, equivalent to signal ``debouncing'' in
engineering, while strong hysteresis leads to bistability, toggling,
and long-term memory~\cite{Lisman:bistability:85,Bialek:memory:01}.
We will look specifically at toggling, because this subsumes all the
other phenomena of interest.

\begin{figure}[ht]
  \begin{center} 
  \includegraphics[scale=0.50]{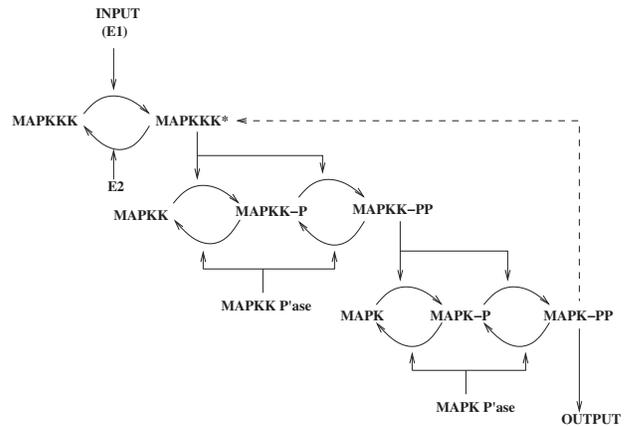}
  \caption{
  The catalytic topology of the mitogen-activated protein kinase
  (MAPK) cascades (from Ref.~\cite{Huang:MAPK:96}, with dashed
  positive feedback arrow representing polyadenylation that inhibits
  degradation of the top protein).  Each level in the diagram
  represents the phosphorylation states of a single protein,
  with phosphorylation and dephosphorylation transitions represented
  by arrows.  Exogenous and internal catalysis is represented by
  (other) arrows pointing to the transition arrows.
    \label{fig:ultrasens} 
  }
  \end{center}
\end{figure}

The molecular mechanisms used repeatedly in signal transduction for
amplification, threshold sensitivity, and switching, are shown in
Fig.~\ref{fig:ultrasens}, as they are instantiated in the
Mitogen-Activated Protein Kinase (MAPK) family of transduction
cascades.  This widely duplicated and diversified homologue family is
used througout the eukaryote kingdom~\cite{Kultz:stress_signals:98},
mostly to regulate gene expression in response to cell-membrane
received signals.  MAPK cascades employ three proteins, each with an
unphosphorylated state, and respectively one, two, and two
phosphorylated states.  Phosphorylation and dephosphorylation of each
protein is catalyzed by exogenous kinases and phosphatases, and in
addition the fully-phosphorylated states of the first two proteins act
as kinases (phosphorylation catalysts) on the proteins following them
in the cascade.  The cascade is thus an actively driven, dissipative
system, maintained away from equilibrium by the supply of activated
(high-energy) phosphate donors.  When used for switching, MAPK
cascades may have positive feedback from the output to the
highest-level protein, either through gene expression or through
inhibition of degradation of the active state~\cite{Huang:MAPK:96}.

\begin{figure}[ht]
  \begin{center} 
  \includegraphics[scale=0.4]{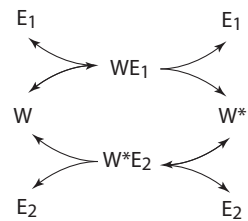}
  \caption{
  The Goldbeter-Koshland minimal phosphorylation couple.  A single
  protein species has an unphosphorylated state $W$ and a
  singly-phosphorylated state $W^{\ast}$.
  Ref.\cite{Goldbeter:hypersensitivity:81} chained such couples to
  model intermolecular phosphorylation, in combination with exogenous
  catalysts that are often present for both transitions as well. 
    \label{fig:goldbeter_koshland} 
  }
  \end{center}
\end{figure}

The structure of MAPK and other cascades was abstracted by Goldbeter
and Koshland~\cite{Goldbeter:hypersensitivity:81} to the minimal
system shown in Fig.~\ref{fig:goldbeter_koshland}, which they propose
as the signaling counterpart to the transistor (a better analogy would
be to the bistable flip-flop, as they use it).  A single protein
species has a single phosphorylation site. Phosphorylation and 
dephosphorylation occur via the action of catalysts/enzymes with 
which the protein can form
enzyme-substrate complexes. Depending on the rate of these reactions, 
the steady state fractions of the phosphorylated (or unphosphorylated)
protein can vary abruptly as a function of these rates.
This analogy has been extended to an elaborate analysis of the
properties~\cite{Lisman:bistability:85} and combinatorial
logic~\cite{Sauro:proteomics:04} of such switches. In particular,
 ~\cite{Lisman:bistability:85} considers the case where the
the phosphorylated
epitope acts as an intermolecular autocatalyst on phosphorylation
transitions of any unphosphorylated proteins in the population, and shows
that this leads to bistability.
However, autocatalytic feedback only creates a bistable 
switch if the response of the
underlying phosphorylation chain is
sigmoidal~\cite{Ferrell:bistability:01}, which in this model requires
saturation of the exogenous phosphatase rate 
via the formation of catalyst-substrate complexes as an intermediate step
between the unphosphorylated and phosphorylated states of the protein.

We note, however, that kinetic control through saturated complex
formation is not the only way to obtain sigmoidal response, because
with {\em two or more} phosphorylation sites per protein, the
concentration of the fully-phosphorylated state is a sigmoidal
function of the ratio of exogenous kinase to phosphatase, even when
all catalysts act in the linear proportional regime (in other words,
when catalyst-substrate complexes act to catalyze transitions
effectively instantaneously, and are limited only by their frequency
of formation through binary encounters). This occurs as long as the
catalytic activity is distributive {\it i.e.}, if at most one
modification (phosphorylation or dephosphorylation) takes place at
each enzyme-substrate encounter~\cite{Gunawardena:switch:05}, 
and ordered (if successive phosphorylations take place at different
residues, an ordered mechanism implies that dephosphorylation takes
place in strictly the inverse order)~\cite{Salazar:phosphate:07}.

Two of the MAPK proteins
have this structure, and more significantly, the intermolecular
catalysis within the cascade is {\em nonspecific} to phosphorylation
reactions on a given protein, though each transition is catalyzed
through an independent event~\cite{Huang:MAPK:96}.  We show below
that, combining this form of sigmoidality with positive feedback, it
is possible to obtain bistability through a {\em non-equilibrium phase
  transition}, in which the individual events of  catalysis
leads to a polarized distribution of phosphorylation 
states of the target protein.
Such population-level cooperative effects, (proposed also in the
context of genetic switches in ~\cite{Sasai:gene_exp:03}), bestow the
stability of macroscopic (thermodynamic) systems on the otherwise
highly stochastic events of phosphorylation and dephosphorylation.  We
suggest that the properties of phase-transition-mediated switching are
one source of adaptive preference for multiple phosphorylation sites
and non-specific catalysis, which one encounters repeatedly (histidine
kinase cascades may have as many as 26 phosphorylation
sites~\cite{Kreegipuu:phosphobase:99}).  

Previous studies have also shown that multisite phosphorylation with
saturation kinetics at each modification step can lead to bistability
even in the absence of feedback
\cite{Markevich:signalling:04,Craciun:bistability:06}. Hence both
kinetic control and population-level polarization can lead robustly to
bistability in some parameter domains.  However, the two mechanisms
are distinguished by their responses to mutations and by their control
parameters.  Single-molecule control causes switching properties to
change if rate kinetics change, in a way that population polarization
does not, while the role of nonspecific catalysis in models with
population-level cooperative effects (at least in the form we will
consider) creates a different kind of sensitivity.  An important
constraint on the evolutionary innovation, preservation, and
diversification of phenotype (any expressed functionality) is the
shape of its neutral
network~\cite{Ancel:mod_RNA:00,Fontana:evo_devo_RNA:02} (the
degenerate space in the genotype/phenotype map with regard to that
functionality).  The phenotype of phase-transition-mediated switching
is more nearly controlled by the topology of the catalytic network
than by its kinetics, an idea that has been proposed as a source of
robustness in the segment-polarity network~\cite{vonDassow:SPN:00},
and theoretically grounded in the case of general enzyme-driven
reaction networks in~\cite{Craciun:bistability:06}.

Note that we do not study spatio-temporal correlations induced by
diffusion of enzymes and/or enzyme inactivation. A recent
study~\cite{Takahashi:spatio-temporal:10} shows that even with the
enzymes acting according to a distributive mechanism, rapid rebindings
of the enzyme molecules to the substrate molecules can lead to a loss
of ultrasenstivity and bistability.  We do not consider the effect of
protein degradation either. Our model is however a first theoretical
fully stochastic study of the MAPK cascade, modelled earlier, to our
knowledge, only via rate equations ~\cite{Huang:MAPK:96,
  Kholodenko:negative:00, Markevich:signalling:04,%
  Salazar:phosphate:07,Heinrich:mathmodels:02} or stochastic
simulations~\cite{Wang:stochastic:06,Kapuy:mol_switches:09}.

In this context, we study quantitatively the three critical properties
of a phase-transition mediated switch: the conditions for existence of
bistability, the noise characteristics of those fluctuations that
preserve the domain in the bistable phase, and the large excursions
that limit memory or reliability of the switch, and which near the
threshold for bistability, can lead to finite-particle number
corrections to that threshold.  These have only been considered
piecemeal before in other models, with conditions for bistability
treated in the infinite-particle (deterministic)
limit~\cite{Tyson:Sniffers:03,Novak:CellCycle:01}, noise from internal
and external sources related through ad hoc response
functions~\cite{Paulsson:noise:01}, and stability treated at the level
of bounds on scaling, for systems already assumed reduced to one
relevant dimension~\cite{Bialek:memory:01}.

\subsection{Reducing to appropriate models}

Most biological literature on this subject focuses on phenomenological
modeling of (usually mean-field behavior in) observed or designed
systems~\cite{vonDassow:SPN:00,Ferrell:xenopus:99,Ferrell:switch:99,%
  Tyson:CellCycle:02,Huang:MAPK:96,Tyson:Sniffers:03,Novak:CellCycle:01,Markevich:signalling:04}. We are however more interested
in the possibility of statistically motivated universality
classification of strategies for switching, which might explain
evolutionary regularities in cascades.  Therefore, in addition to
idealizing molecular mechanisms responsible for sigmoidal response and
positive feedback as properties of single protein species, to make the
polarization-based equivalent of the flip-flop from adding feedback to
our Fig.~\ref{fig:goldbeter_koshland} (equivalent to Fig.~12 of
Ref.~\cite{Sauro:proteomics:04}), we advisedly exploit symmetry, of
either the catalytic topology or the parameters, to make analysis
tractable. This approach also aids in decomposing effects responsible
for switching, and relating these to other equilibrium or
non-equilibrium phase transitions.  Thus our minimal models
deliberately differ from the familiar cascade families in areas not
directly related to the production of
switching~\cite{Krishnamurthy:Signaling:07}.  The model of Markevich
{\it et al} ~\cite{Markevich:signalling:04} is also in this category
in demonstrating bistability (via kinetic control) at the level of a
single stage of the MAPK cascade.

\begin{figure}[ht]
  \begin{center} 
  \includegraphics[scale=0.45]{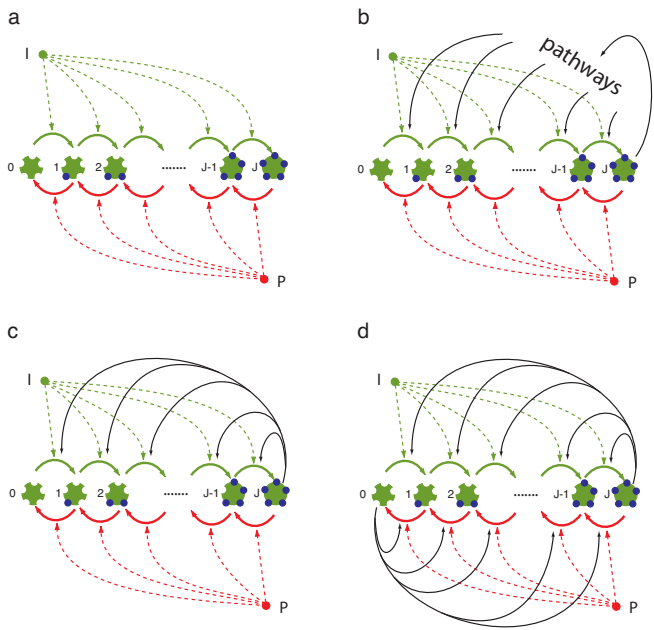}
  \caption{
    Multisite phosphorylation and feedback structure.  (a) This panel
    depicts the basic phosphorylation chain without feedback in which
    a target protein with $J$ sites is phosphorylated by a kinase $I$
    and dephosphorylated by a phosphatase $P$. The ordered succession
    of phosphorylations yields $J+1$ modification states, labelled
    $0$, $1$, $\ldots$, $J$. (b) The fully phosphorylated target
    protein relays a signal into pathways that eventually feed back on
    the phosphorylation chain. (c) Simplification of (b) in which the
    fully phosphorylated target protein acquires kinase activity and
    directly feeds back on the chain. We refer to this network
    configuration as the asymmetric circuit. (d) Schematic of the
    network with symmetric feedback in which the substrate protein is
    bifunctional, whereby the fully (de)phosphorylated form catalyzes
    (de)phosphorylation of its own precursors.
    \label{fig:schemes} 
  }
  \end{center}
\end{figure}

The model we propose for a cooperative-phosphorylation switch is shown
in Fig.~\ref{fig:schemes}.  Each of $N$ molecules of a single type of
protein has $J$ phosphorylation sites indexed $j \in 1 , \ldots , J$,
which we suppose for simplicity to be phosphorylated and
dephosphorylated in a definite order. All phosphorylations are
catalyzed by exogenous kinases, and all dephosphorylations by
exogenous phosphatases.  Because the enzymes are assumed to operate in
the linear regime where complex formation is not rate limiting, the
catalytic rate per reaction is proportional to the numbers $I$ and $P$
of kinase and phosphatase particles respectively (We set the constant
of proportionality equal to one by choice of the units of time).
The site modifications occur in a specific order, thus sidestepping
combinatorial complexity. 
Furthermore, phosphorylation and dephosphorylation of substrate
proteins is assumed to follow a distributive mechanism, whereby a
kinase (phosphatase) enzyme dissociates from its substrate between
subsequent modification events \cite{Ferrell:MAPK:97, Burack:nonprocessive:97}. 
Hence the substrate has $J+1$ states.

We obtain positive feedback from intermolecular autocatalysis;
specifically, proteins in the $j = J$ state are kinases that act
interchangeably with the exogenous kinases (unequal catalytic power
between $j=J$ and exogenous kinases can easily be added, at the cost
of another parameter).  The phosphorylation chain with feedback is shown 
in the bottom half of Fig.~\ref{fig:schemes}. Panel (c) depicts an 
asymmetric topology in which the fully phosphorylated substrate catalyzes 
its own phosphorylation, while panel (d) shows the symmetric version 
in which a substrate molecule is bifunctional, acting as both a 
kinase and phosphatase depending on its modification state. Kinase $I$ 
and phosphatase $P$ are exogenous forces on the modification of the 
substrate, but the feedback is an endogenous force whose strength 
is proportional to the occupancy of the end-states of the chain. 
This occupancy is subject to intrinsic fluctuations and depends on 
the total number of substrate molecules.

 The assumption of intermolecular autocatalysis is
standard~\cite{Lisman:bistability:85}, and we consider below the
self-consistent backgrounds with kinase-only autocatalysis, as the
nearest equivalent to the kinetically-controlled
switch~\cite{Goldbeter:hypersensitivity:81}. In order to analyse 
the model beyond mean-field theory (using field
theoretic techniques) we go further in the interest of simplicity, and
symmetrize the topology as in Fig.~\ref{fig:schemes}d, by making the
unphosphorylated state (indexed $j=0$) a phosphatase, interchangeable
with the exogenous phosphatases. 

The feedback topology of the model caricatures a few elements present
in biological systems. One such element is the competition between
antagonistic pathways that may underlie cellular decision processes
(for example \cite{Gaudet:compendium:05}). A multisite phosphorylation
chain of the type considered here could function as an evaluation
point between competing and antagonistic pathways influenced by
different active phosphoforms of the chain, provided these pathways
feed back to the chain. In a less extreme case, the fully
phosphorylated form activates another kinase which then interacts with
the chain. In these scenarios, feedback is mediated by a series of
intervening processes, which may well affect the propagation of
fluctuations. Yet, if delays are not too large, the collapsed scheme
of Fig.~\ref{fig:schemes}c could be a reasonable proxy with the added
benefit of mathematical tractability.

A scenario corresponding more literally to our model involves a
bifunctional substrate capable of both kinase and phosphatase
activity, depending on the substrate's modification state. One example
is the HPr kinase/P-Ser-HPr phosphatase (HprK/P) protein, which
operates in the phosphoenolpyruvate:carbohydrate phosphotransferase
system of gram-positive bacteria. Upon stimulation by
fructose-1,6-bisphosphate, HprK/P catalyzes the phosphorylation of HPr
at a seryl residue, while inorganic phosphate stimulates the opposing
activity of dephosphorylating the seryl-phosphorylated HPr (P-Ser-HPr)
\cite{Kravanja:hprK:99}. Another example of a bifunctional
kinase/phosphatase is the NRII (Nitrogen Regulator II) protein. It
phosphorylates and dephosphorylates NRI. NRI and NRII constitute a
bacterial two-component signaling system, in which NRII is the
``transmitter" and NRI the ``receiver" that controls gene
expression. NRII autophosphorylates at a histidine residue and
transfers that phosphoryl group to NRI. The phosphatase activity of
NRII is stimulated by the PII signaling protein (which also inhibits
the kinase activity). Several other transmitters in bacterial
two-component systems seem to possess bifunctional kinase/phosphatase
activity \cite{Ninfa:protein:91}.


  %
  %

Both the network with asymmetric topology (auto-kinase only,
Fig.~\ref{fig:schemes}c) and the network with symmetric topology
(Fig.~\ref{fig:schemes}d) but asymmetric catalytic concentrations $I
\neq P$ undergo formally first-order phase transitions, so that
regions of bistability are always metastable at finite $N$.  However,
in the topologically symmetric case, these continue smoothly through a
second-order transition at $I=P$, in which symmetry of both topology
and parameters ensures exact bistability with finite residence time in
domains, at all $N$ where the phase transitions exists.  This
simplification permits us to estimate the residence times with an
expansion in semiclassical stationary points of an effective action,
without encountering the complexities of path integrals for metastable
processes~\cite{Coleman:AoS:85}, though numerically we expect this
also to be a good approximation to residence times in metastable
states with similar ``barrier heights'' in the first-order case.
Symmetry also permits the closed-form computation of the noise kernel
about the monostable phase with a unique equilibrium, which generates
a natural measure for ``weakness'' or ``strongness'' of the
first-order transitions at nearby values of $I/P$ as a function of
$N$.  We therefore perform a thorough analysis of the second-order
transition, to establish methods and provide a reference solution to
qualitatively understand the mechanisms of bistability and
metastability in the more general cases with similar stochastic
structure.

\subsection{Methods of treatment for the stochastic problem}

While differential equations for mean chemical concentrations (the
current standard method of analysis) can give good estimates of the
existence of hysteresis and bistability when approximating systems
with as few as tens to hundreds of molecules, they of course preclude
the treatment of noise, fluctuation-induced corrections to mean-field
behavior at small particle number or near critical points, and large
excursions such as domain flips (when the system switches from one
bistable state to another).  Pure mass-action models also ignore
spatial constraints such as scaffolding by the cytoskeleton or the
proteins themselves, and the dimensionality of physical diffusion in
the cytosol or membranes.

A better approximation is given by the master equation for the
probability of instantaneous particle distributions in models like
that of Fig.~\ref{fig:schemes}, which in principle captures all orders
of stochastic processes, though such simple models still omit spatial
effects.  The general properties of the master-equation (in the
diffusion, or ``Fokker-Planck'' approximation) for a one-dimensional
switch have been used to obtain scaling relations and loose bounds on
the stability achievable from such a switch as a function of the
number of molecules it employs~\cite{Bialek:memory:01}.

Operator methods, analogous to Hamiltonian methods in quantum
mechanics, have been developed in reaction-diffusion
theory (see ~\cite{Mattis:RDQFT:98} for a review) 
to efficiently handle the collective
excitations that diagonalize general master equations without
time-reversal symmetry.  These have been used in the context of gene
expression~\cite{Sasai:gene_exp:03} to estimate the number of stable
cell types made possible by many randomly combined transcription
factors, making use of similarities to ground states of random-bond
Ising models. 

From the operator-valued evolution kernel, one can obtain an
equivalent path-integral representation by expanding at each time in a
basis of coherent states~\cite{Eyink:action:96,Cardy:FTNEqSM:99}.
Stationary-field expansion in the path integral generalizes the
classical differential equation for concentrations to consistently
incorporate fluctuation effects (by means of a
perturbatively-corrected effective action~\cite{Smith:LDP_SEA:11}),
and the sum over ``approximate stationary points'' of locally
least-action identify the typical configuration histories associated
with domain flips.  More sophisticated approaches, similar to those
used here, have also been used to incorporate fluctuation effects into
efficient lumped-parameter expansions for networks with multiple
timescales~\cite{Sinitsyn:CG_chem_nets:09}. 

Master equations can also be solved numerically by the Gillespie
algorithm~\cite{Gillespie:QTMP:94}, or simulated directly, and we use
such simulations to validate our anlaytic results below.  The lack of
convenient symmetries in real biomolecular systems promises to make
analysis intractable for most quantitative phenomenology, and recourse
to numerics is likely to be the only general-purpose solution.
However, the path integral's separation of moments in a natural
small-parameter expansion, and of perturbative noise from formally
non-perturbative large excursions, provides an intuitive decomposition
of the mechanisms fundamental to switching and stability.  At
mean-field approximation, we find surprising similarities of the phase
transition in this driven system to the magnetization transition in
the discrete, equilibrium, mean-field Ising ferromagnet, and a
transition between this classical critical behavior at finite $J$ and
a condensation effect more similar to Bose-Einstein condensation at $J
\rightarrow \infty$.  The algebraic distinction between
self-consistent backgrounds, perturbative fluctuations, and
non-perturbative domain flips, elementary in the analysis, is also a
subtle distinction, difficult to make without systematic measurement
biases, in the numerics.

\subsection{Main results from the analytical treatment}

The mean-field results, which are reported in detail in ~\cite{Krishnamurthy:Signaling:07}, and which can also be recovered from our effective-action treatment in this paper, reproduce 
the standard differential equations for mass action.  The stationary
states arise from conditions of detailed balance between
phosphorylation and dephosphorylation, self-consistent with the
concentrations they produce of autocatalytic phosphoepitopes in
relation to exogenous catalysts.  Specifically, we show how both
symmetric and asymmetric topologies create domains of mono- and
bi-stability in the parameter space $\left( I/N, P/N \right)$, and how
the population asymmetry in the self-consistent state depends on the
coupling $g \equiv N / \sqrt{IP}$ and exogenous asymmetry $I/P$.

The perturbative expansion in Gaussian fluctuations about the
self-consistent background provides a systematic construction of the
noise spectrum of the phosphorylation chain.  At lowest order it
predicts a cusp $\sim 1 / \left| g - g_c \right|$ in the variance of
the order parameter, equivalent to the Curie-Weiss prediction for the
spin-$1/2$ mean-field ferromagnet.  More surprising, we find that the
entire perturbative approximation to the noise spectrum on all sites
is generated from a single bare mode, effectively coupled to a single
Langevin field.  This result replaces the {\em ad hoc} noise kernels
one must entertain in the absence of a first-principles
treatment~\cite{Paulsson:noise:01,Aurell:epigenetics:02}.

The nonperturbative expansion in semiclassical configurations of
locally least action predicts the leading large-$N$ dependence of the
domain residence time in the bistable regime, as a function of the
dimensionless rates of the problem $I/P$ and $\sqrt{IP}/N$ (though
here we solve only for the symmetric case $I/P = 1$, where bistability
remains exact at finite particle numbers).  These configurations, the
dissipative equivalent to the instantons of Euclidean equilibrium
field theory~\cite{Coleman:AoS:85}, solve two problems.  First, from
the high-dimensional configuration space of the $N$-particle, $\left(
J+1 \right)$-site chain, they extract the one-dimensional contour of
most likely configurations to mediate domain flips, assumed given in
Ref.~\cite{Bialek:memory:01}.  Second, the action along this
trajectory, $\propto N$ at fixed $\left( I/N , P/N \right)$, is the
leading exponential in the residence time, for which
Ref.~\cite{Bialek:memory:01} correctly predicts the scaling but gives
no algorithm to compute the coefficient (known in large-deviations
literature as the \emph{rate function}~\cite{Touchette:large_dev:09}).

Similar leading-exponential dependencies have been computed in
Ref's.~\cite{Aurell:epigenetics:02,Roma:epigenetics:05}.  For
reference to this work, we note that the passage to the diffusion
limit or Fokker-Planck equation in Ref.~\cite{Bialek:memory:01}, and
the closely-related use of the Gaussian approximation for fluctuations
in Ref.~\cite{Aurell:epigenetics:02},\footnote{This form is originally
  due to Onsager and Machlup~\cite{Onsager:Machlup:53}.} are formally
uncontrolled approximations, whose limitations and ranges of validity
are pointed out in Ref.~\cite{Roma:epigenetics:05}.  One purpose for
our paper is to present the larger systematic analysis within which
such approximations arise.

\subsection{Layout of the paper}

Sec.~\ref{sec:detailed_balance} introduces the master equation for the
model class of Fig.~\ref{fig:schemes} c and d, and derives the phase
diagram for steady states from conditions of detailed balance of the
mean particle numbers.  Sec.~\ref{sec:Op_State_PI} converts the master
equation, first into the equivalent representation in terms of a state
in a Hilbert space, and then into the equivalent path-integral
representation through an expansion in intermediate Poisson
distributions.  Sec.~\ref{sec:flucts} derives the perturbative
expansion in fluctuations about the mean fields of the path integral,
including the equivalent representation in terms of a Langevin
equation, and the leading-order perturbative approximation to the
fluctuations in the order parameter.  Sec.~\ref{sec:large_dev} then
considers the enlarged expansion in approximate stationary points
needed to derive the trajectories and rate of domain flips.  Finally,
Sec.~\ref{sec:conclusions} summarizes the consequences of these
technical results for the conceptual understanding of biomolecular
signal transduction and switching.

\section{Master equation and mean-field backgrounds}
\label{sec:detailed_balance}

An instantaneous configuration of $N$ proteins on the $J+1$ sites of
Fig.~\ref{fig:schemes} defines a vector $n \equiv \left( n_0 ,
n_1 , \ldots , n_J \right)$, where $n_j$ is the number on site $j$.
Fixed particle number implies that $n$ lives on the integer lattice in
the $J$-simplex $\sum_{j = 0}^{J} n_j \equiv N$.  We denote a
(generally time-dependent) probability distribution on configurations
$P \! \left( n \right)$, and suppress the time index $t$ in the
notation.

A stochastic process for particle hopping is completely defined by the
master equation for $P \! \left( n \right)$, which is the ``probability
inheritance'' equation induced by the transition probabilities on the
simplex.  For Fig.~\ref{fig:schemes} with catalytic rates
proportional to the number of catalytic particles, this is 
\begin{widetext}
\begin{eqnarray}
  \frac{
    \partial 
  }{
    \partial t
  } 
  P \! \left( n \right) 
& = & 
  \sum_{j = 0}^{J-1}
  \bigl[
    \left(
      I + n_J - {\delta}_{J , j+1}
    \right)
    \left(
      n_j + 1
    \right)
    P \! \left( n + 1_j - 1_{j+1} \right) - 
    \left(
      I + n_J 
    \right)
    n_j P \! \left( n \right)
  \bigr.
\nonumber \\
& & 
  \bigl.
    \mbox{} + 
    \left(
      P + n_0 - {\delta}_{0 , j}
    \right)
    \left(
      n_{j+1} + 1
    \right)
    P \! \left( n - 1_j + 1_{j+1} \right) - 
    \left(
      P + n_0 
    \right)
    n_{j+1} P \! \left( n \right)
  \bigr] , 
\label{eq:master_eqn}
\end{eqnarray}
\end{widetext}
where $1_j$ denotes the vector with $j$th 
component equal to $1$ and all other components zero.

Our assumption that phosphorylation and dephosphorylation happen in a
definite sequence makes transition rates from site $j$ proportional to
$n_j$ and the catalyst concentration, without additional combinatorial
factors.  For the asymmetric (auto-kinase only) topology, the factors
$n_0$ and ${\delta}_{0 , j}$ in the second line of
Eq.~(\ref{eq:master_eqn}) are absent, and dephosphorylation depends
only on the $n_j$ and the exogenous phosphatase number $P$.

Time-dependent average particle numbers on each site are defined as 
\begin{equation}
  \left<
    n_j
  \right> \equiv 
  \sum_{n}
  P \! \left( n \right)
  n_j , 
\label{eq:part_aves_def}
\end{equation}
and it is easy to see from Eq.~(\ref{eq:master_eqn}) that
$\sum_{j=0}^J d \left< n_j \right> / dt \equiv 0$.  

It is also useful to write the equation for the center-of-mass of the system
defined as $C \equiv \sum_{n,j}jn_jP(n) $. This becomes
\begin{eqnarray}
  \frac{
    \partial 
  }{
    \partial t
  } C &=&
N \left( I-P \right)
+
\left[P\left<n_0\right> - I \left<n_J\right>\right] 
\nonumber \\
&+& N \left[\left<n_J\right> - \left<n_0\right>\right]
+ \left[\left<n_0^2\right> - \left<n_J^2\right>\right]
\label{eq:cm}
\end{eqnarray}
As we will see later, this exact equation can be used to estimate
the fluctuations of the order parameter for large $J$.

In what follows, we first look at the mean-field approximation already
elaborated on in \cite{Krishnamurthy:Signaling:07}. The mean-field
approximation for evolution of $\left< n_j \right>$ under
Eq.~(\ref{eq:master_eqn}) replaces all joint expectations with
products of marginals: $\left< n_j n_J \right> \approx \left< n_j
\right> \left< n_J \right>$, etc.

\subsection{Detailed balance: symmetric topology}
\label{subsec:det_bal_symm}

Under the mean-field approximation, the system of N interacting particles
essentially decouples into a system of N independent particles
executing a random walk on the lattice of $J+1$ sites. Within this
approximation detailed balance of
phosphorylation and dephosphorylation between adjacent sites in the
chain holds, and  
depends on a catalytic ratio which we will denote $x$.  For the
symmetric topology, $ x \equiv \left( I + \left< n_J \right> \right) /
\left( P + \left< n_0 \right> \right) $, and we recognize two
convenient nondimensional parameters:
\begin{equation}
  \frac{I}{P} \equiv 
  e^{\lambda}, 
\label{eq:IP_lambda_def}
\end{equation}
and 
\begin{equation}
  \frac{N}{\sqrt{IP}} \equiv
  g . 
\label{eq:g_introd}
\end{equation}
As autocatalysis, scaled by $N$, induces bistable order ({\it i.e.} it
favours configurations in which most particles are piled up towards
one or the other end of the chain), and exogenous catalysis, scaled by
$I$ and $P$, induces homogeneity (configurations in which particles
are uniformly spread out on the chain), $g$ is the coupling strength
of the model, with strong coupling favoring broken symmetry.  $I/P$ is
then the measure of exogenous asymmetry.

In terms of these and the fractional occupations $\left< n_j \right> /
N$, the catalytic ratio may be written
\begin{equation}
  x \equiv 
  \frac{
    e^{\lambda/2} + g \left< n_J \right> / N
  }{
    e^{-\lambda/2} + g \left< n_0 \right> / N
  } \equiv
  e^{\xi} . 
\label{eq:x_def_gen_symtop}
\end{equation}
By induction on $j$, time-independent solutions satisfy 
\begin{equation}
   \left< n_j \right> = 
  x^j
  \left< n_0 \right> , \;  
  0 \le j \le J , 
\label{eq:0_J_x_reln}
\end{equation}
and the normalization 
\begin{equation}
  N =
  \sum_{j=0}^J 
  \left< n_j \right> = 
  \left< n_0 \right>
  \frac{
    1 - x^{J+1}
  }{
    1 - x
  } . 
\label{eq:tot_num_const}
\end{equation}
From Eq.~(\ref{eq:x_def_gen_symtop}) and Eq.~(\ref{eq:0_J_x_reln}) we
can evaluate 
\begin{equation}
  \frac{g}{N}
  \left< 
    n_J - n_0
  \right> = 
  \frac{
    2 
    \sinh
    \left(
      \frac{\xi - \lambda}{2}
    \right) 
    \sinh
    \left(
      \frac{J}{2} \xi
    \right) 
  }{
    \sinh
    \left(
      \frac{J - 1}{2} \xi
    \right) 
  } , 
\label{eq:self_con_ord_parm_asy}
\end{equation}
and we can rewrite Eq.~(\ref{eq:tot_num_const}) as 
\begin{equation}
  \frac{
    \left< 
      n_J - n_0
    \right> 
  }{
    N 
  } = 
  \frac{
    2 
    \sinh
    \left(
      \frac{\xi}{2}
    \right) 
    \sinh
    \left(
      \frac{J}{2} \xi
    \right) 
  }{
    \sinh
    \left(
      \frac{J + 1}{2} \xi
    \right) 
  } . 
\label{eq:self_con_num_asy}
\end{equation}

When Eq.~(\ref{eq:self_con_ord_parm_asy}) and
Eq.~(\ref{eq:self_con_num_asy}) are nonzero, they have the ratio
\begin{equation}
  g  = 
  \frac{
    \sinh
    \left(
      \frac{\xi - \lambda}{2}
    \right) 
    \sinh
    \left(
      \frac{J + 1}{2} \xi
    \right) 
  }{
    \sinh
    \left(
      \frac{\xi}{2}
    \right) 
    \sinh
    \left(
      \frac{J - 1}{2} \xi
    \right) 
  } . 
\label{eq:g_from_xi_asy}
\end{equation}

\begin{figure}[ht]
  \begin{center} 
  \includegraphics[scale=0.5]{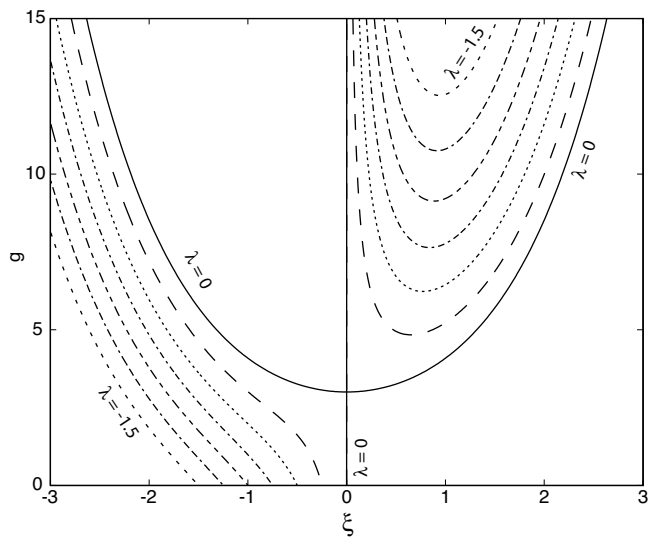}
  \caption{
  $g \! \left( \xi \right)$ from Eq.~(\ref{eq:g_from_xi_asy}) along
    contours of constant $\lambda$, for which the two branches proceed
    outward from the $\xi=0$ line in order of increasing
    $\left| \lambda \right|$ (different $\lambda$ values are depicted 
    by different line styles). The branches with the lowest and largest 
    values of $\lambda$ are marked on the figure.   The other 
    $\lambda$ values can be read on the
    left from the intersection with the abscissa where $\xi =
    \lambda$. Where $g \!  \left( \xi \right)$ has a single-valued
    inverse, that function $\xi \!  \left( g \right)$ defines the
    unique steady-state distributions.  Where the inverse is
    triple-valued, the largest $\left| \xi \right|$ are stable
    solutions, and the central branch is unstable.
    \label{fig:g_asy_Jis2_neg_l} 
  }
  \end{center}
\end{figure}

For $I \neq P$, Eq.~(\ref{eq:g_from_xi_asy}) always holds (though it
may be negative or singular), while for $I = P$ we have the
possibility of the degenerate case where $\left< n_J - n_0 \right> =
0$ and $g$ is unconstrained.  For $g < g_C \! \left( \lambda = 0
\right)$ (a second-order critical point) this is the stable asymptotic
distribution, while for $g > g_C \! \left( \lambda = 0 \right)$ it is
unstable.  The graph of $g$ versus $\xi$ for a few (non-positive)
values of $\lambda$ at $J=2$ is shown in
Fig.~\ref{fig:g_asy_Jis2_neg_l}.  (Positive $\lambda$ generate curves
reflected through $\xi = 0$.)  The graph defines a pseudo-inverse $\xi
\! \left( g \right)$, which gives the stationary solutions within the
mean-field approxiation.  Where $\xi$ is triple-valued (not a
well-defined inverse), the central branch is in all cases unstable,
and the two outer branches are stable.

The character of the curves in Fig.~\ref{fig:g_asy_Jis2_neg_l} is
preserved for all $J > 2$, though the derivative of the stable curve
for $I = P$ above its critical point becomes discontinuous at $\xi=0$
for $J \rightarrow \infty$.  This discontinuity is related to the
transition from Curie-Weiss to Bose-Einstein-like behavior of the
order parameter, discussed below.  The curves corresponding to all
$\lambda$ have regular limits at large $J$.

We can identify a set of $g_C \! \left( \lambda \right)$ as the local
minima in Fig.~\ref{fig:g_asy_Jis2_neg_l} above which the $\xi \!
\left( g \right)$ graph becomes triple-valued.  The $\lambda
\rightarrow 0$ limit of these minima smoothly converges on the
second-order critical coupling
\begin{equation}
  g_C \! \left( \lambda = 0 \right) \equiv
  \frac{
    J+1
  }{
    J-1
  } . 
\label{eq:g_crit_def}
\end{equation}

\begin{figure}[ht]
  \begin{center} 
  \includegraphics[scale=0.5]{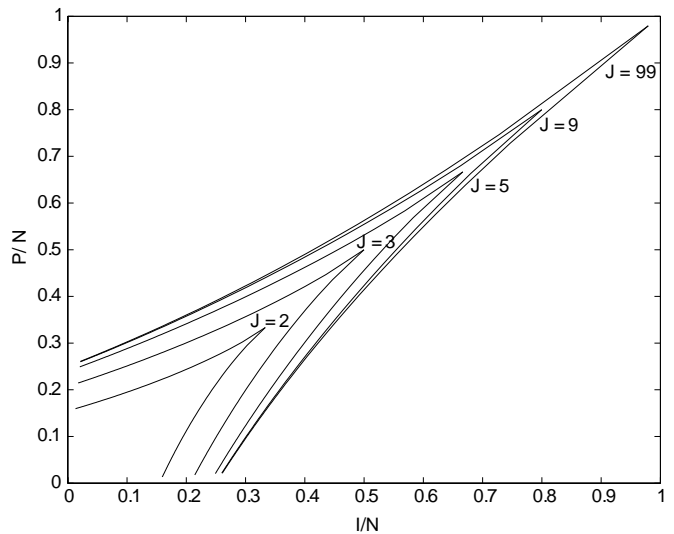}
  \caption{
  Phase diagram for the symmetric topology, from the minima $g_C \!
  \left( \lambda \right)$ of Fig.~\ref{fig:g_asy_Jis2_neg_l}, and
  their equivalents for a range of $J$.  The regions inside the
  chevrons are bistable.  
    \label{fig:sym_phase_diag} 
  }
  \end{center}
\end{figure}

Converting the pair $\bigl( \lambda, g_C \! \left( \lambda \right)
\bigr)$ to $\left( I/N , P/N \right)$ values yields the phase diagram
shown in Fig.~\ref{fig:sym_phase_diag} for a range of $J$ values.  The
interior region $I \sim P$ and sufficiently small $\sqrt{IP} / N
\equiv 1/g$ is bistable, and outside this region the sign of $\xi =
\log x$ equals that of $\lambda \equiv \log I/P$. As we demonstrate in
\cite{Krishnamurthy:Signaling:07}, these theoretical estimates match
very well with data from Monte carlo simulations.

\subsection{Detailed balance: asymmetric topology}

In the auto-kinase-only asymmetric model, positive $\lambda$ ($I > P$)
is never bistable, because the particles are already biased toward
$n_J$, the only site with positive feedback.  Therefore we graph only
$\lambda \le 0$, though the algebraic solutions are valid everywhere. 

Instead of Eq.~(\ref{eq:x_def_gen_symtop}), the catalytic ratio is 
$ x \equiv 
  \left( I + \left< n_J \right> \right) / P 
$, 
which reduces in nondimensional parameters to 
\begin{equation}
  x \equiv 
  \frac{
    e^{\lambda/2} + g \left< n_J \right> / N
  }{
    e^{-\lambda/2} 
  } \equiv
  e^{\xi} . 
\label{eq:x_def_gen_asytop}
\end{equation}
Equations~(\ref{eq:0_J_x_reln}) and~(\ref{eq:tot_num_const}) still
hold, but instead of Eq.~(\ref{eq:self_con_ord_parm_asy}) we choose
the reduction
\begin{equation}
  g
  \frac{
    \left< 
      n_0
    \right> 
  }{
    N 
  } = 
  \frac{
    2 
    \sinh
    \left(
      \frac{\xi - \lambda}{2}
    \right) 
  }{
    \exp
    \left[
      \left( J - \frac{1}{2} \right) \xi
    \right] 
  } . 
\label{eq:self_con_ord_parm_asytop}
\end{equation}
The appropriate reduction of Eq.~(\ref{eq:tot_num_const}), counterpart
to Eq.~(\ref{eq:self_con_num_asy}), is now 
\begin{equation}
  \frac{
    \left< 
      n_0
    \right> 
  }{
    N 
  } = 
  \frac{
    \sinh
    \left(
      \frac{\xi}{2}
    \right) 
  }{
    \exp
    \left(
      \frac{J}{2} \xi
    \right) 
    \sinh
    \left(
      \frac{J + 1}{2} \xi
    \right) 
  } . 
\label{eq:self_con_num_asytop}
\end{equation}
Eq.~(\ref{eq:self_con_ord_parm_asytop}) and
Eq.~(\ref{eq:self_con_num_asytop}) are regular at all $\xi$, so we
always have a defined function $g \! \left( \xi \right)$, of the form
\begin{equation}
  g  = 
  \frac{
    2 
    \sinh
    \left(
      \frac{\xi - \lambda}{2}
    \right) 
    \sinh
    \left(
      \frac{J + 1}{2} \xi
    \right) 
  }{
    \sinh
    \left(
      \frac{\xi}{2}
    \right) 
    \exp
    \left(
      \frac{J - 1}{2} \xi
    \right) 
  } . 
\label{eq:g_from_xi_asytop}
\end{equation}

\begin{figure}[ht]
  \begin{center} 
  \includegraphics[scale=0.5]{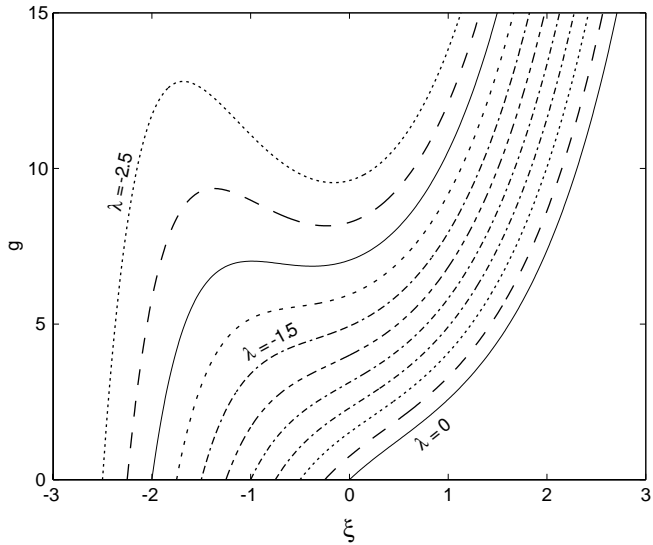}
  \caption{
  $g \! \left( \xi \right)$ along contours of constant $\lambda$ for
  the asymmetric topology, $J=2$ and the same $\lambda$ values as in
  Fig.~\ref{fig:g_asy_Jis2_neg_l}. The maximal $\lambda$ for
  bistability generates the curve whose minimum derivative is zero. 
    \label{fig:g_top_asy_Jis2} 
  }
  \end{center}
\end{figure}

A graph at $J=2$, which is the asymmetric-topology counterpart to
Fig.~\ref{fig:g_asy_Jis2_neg_l}, is shown in
Fig.~\ref{fig:g_top_asy_Jis2}.  In the bistable phase, there are still
three branches for $\xi$ at given $g$, with the outer two stable and
the central one unstable.  The obvious differences are that now there
is a maximal $\lambda$ for bistability, and that the leftmost stable
branch at any $\lambda$ moves positively in $\xi$ as $g$ increases
because of the asymmetric topology, whereas in the symmetric topology
it moved negatively in $\xi$.

\begin{figure}[ht]
  \begin{center} 
  \includegraphics[scale=0.5]{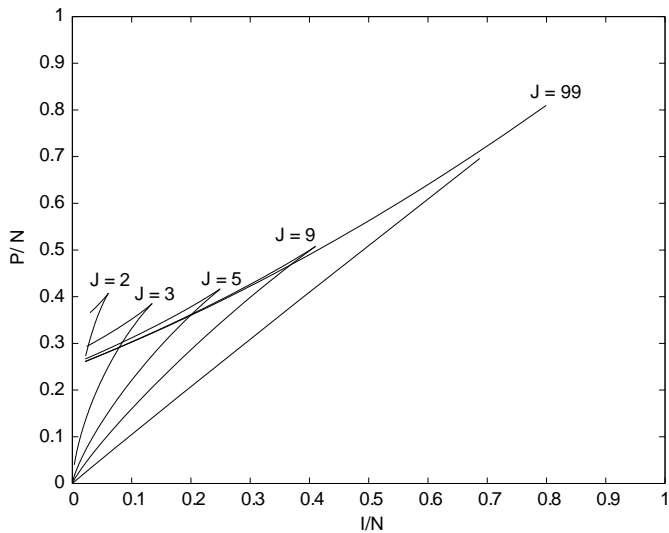}
  \caption{
  The phase diagram for auto-kinase feedback only.  The domains of
  bistability are somewhat smaller than in
  Fig.~\ref{fig:sym_phase_diag}, and are shifted toward smaller $I/P$,
  but otherwise the characteristics are qualitatively and even
  quantitatively similar.
    \label{fig:asym_phase_diag} 
  }
  \end{center}
\end{figure}

At any $\lambda$ below a (negative) $J$-dependent threshold, we can
extract the minimal and maximal $g$ values for bistability (below the
minimum, $\xi$ follows $\lambda \equiv \log \left( I/P \right)$
qualitatively; above the maximum, only the largest-$\xi$ branch is
stable because of too-strong positive feedback).  Inverting
$g_{\mbox{\scriptsize min}} \! \left( \lambda \right)$ and
$g_{\mbox{\scriptsize max}} \! \left( \lambda \right)$ to $\left( I/N
, P/N \right)$, we obtain the phase diagram for bistability shown in
Fig.~\ref{fig:asym_phase_diag}.

The upper boundary of each bistable region, defined by
$g_{\mbox{\scriptsize min}} \! \left( \lambda \right)$, is a distorted
counterpart to the upper $g_C \! \left( \lambda \right)$ branch in
Fig.~\ref{fig:sym_phase_diag}, and the two converge to the same limit
as $J \rightarrow \infty$ (where feedback from $n_0$ becomes
irrelevant).  The lower boundary, defined by $g_{\mbox{\scriptsize
max}} \! \left( \lambda \right)$, replaces the reflected lower $g_C \!
\left( \lambda \right)$ branch in Fig.~\ref{fig:sym_phase_diag}, and
converges to the diagonal $I = P$ at $J \rightarrow \infty$.

Thus we see that classically, the first-order phase transitions are
similar for symmetric and asymmetric feedback topology, one being
deformable into the other in the $\left( I/N , P/N \right)$ parameter
space.  (We could have performed this continuation smoothly by
weighing the $n_0$ catalytic strength with a parameter $\varepsilon
\in \left[ 0, 1 \right]$.)  Further, the first-order transition in $g$
along any $I/P$ ray in the symmetric topology continues smoothly
through the second-order transition at $I/P = 1$, at the apex of the
domain of bistability.  Whether the first-order transitions in the
neigborhood of $I/P = 1$ are strong or weak depends on the
$\xi$-support of the stationary solution for $P \! \left( n \right)$
(a function of $N$), in relation to the difference between the stable
mean values at $g_C \!  \left( \lambda \right)$.

\subsection{Phase transition and order parameter versus $J$}

\begin{figure}[ht]
  \begin{center} 
  \includegraphics[scale=0.475]{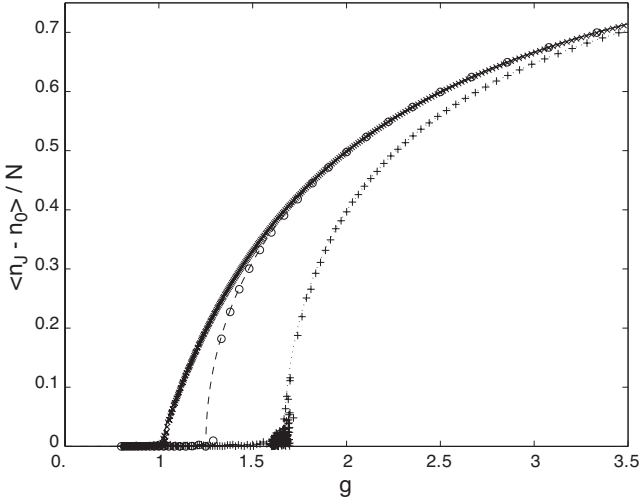}
  \caption{
  Order parameter $\left| \left< n_J - n_0 \right> \right| / N$ for
  the symmetric phosphorylation chain: mean-field theory (lines) and
  simulations (symbols).  $J+1 = \left[ 5, 10, 100 \right]$
  corresponds to $\left[ \mbox{dot}, \mbox{dash}, \mbox{solid}
  \right]$ for lines and $\left[ \mbox{+}, \mbox{o}, \mbox{$\times$}
  \right]$ for symbols. Particle numbers used in the simulations are
  respectively $N = [4000, 2000, 400]$.
    \label{fig:MFT_ord_parms_data_paper} 
  }
  \end{center}
\end{figure}

We now restrict attention to the case of symmetric topology and
exogenous catalysis setting $I = P \equiv q$, and consider the
behavior of the natural mean-field order parameter $\left| \left< n_J
- n_0 \right> \right| / N$ as a function of $J$.  Expanding
Eq.~(\ref{eq:self_con_num_asy}) in small $\xi$, and inverting
Eq.~(\ref{eq:g_from_xi_asy}) relative to $g_c \equiv g_C \! \left(
\lambda = 0 \right)$ in Eq.~(\ref{eq:g_crit_def}), we find the
mean-field critical scaling of the discrete Ising ferromagnet, up to a
$J$-dependent prefactor:
\begin{equation}
  \frac{
    \left| 
      \left< 
        n_J - n_0
      \right>
    \right|
  }{
    N 
  } \approx
  \frac{
    \sqrt{6 J}
  }{
    J + 1
  }
  {
    \left( 
      \frac{g}{g_c} - 1
    \right)
  }^{1/2} . 
\label{eq:ord_p_scaling_CW}
\end{equation}
The small-$\xi$ approximation is valid for $g - g_c \lesssim g_c - 1$,
above which the order parameter saturates to a $J$-independent
envelope value
\begin{equation}
  \frac{
    \left| 
      \left< 
        n_J - n_0
      \right>
    \right|
  }{
    N 
  } \rightarrow
  1 - 
  \frac{
    1 
  }{
    g 
  } . 
\label{eq:ord_p_scaling_BE}
\end{equation}
The exact mean-field prediction for $\left| \left< n_J - n_0 \right>
\right| / N$ versus $g$ from Equations~(\ref{eq:self_con_num_asy})
and~(\ref{eq:g_from_xi_asy}) is compared to numerical simulations for
$J+1 = \left[ 5, 10, 100 \right]$, in
Fig.~\ref{fig:MFT_ord_parms_data_paper}.

Since $g_c - 1 \rightarrow 2 / J$ for large $J$,
Eq.~(\ref{eq:ord_p_scaling_BE}) also gives the behavior in the formal
$J \rightarrow \infty$ limit.  The derivative of the order parameter
converges to one in arbitrarily small neighborhoods of the critical
point, rather than to $\infty$ as in the Curie-Weiss regime; thus $J
\rightarrow \infty$ defines a different universality class than any
finite $J$.  Qualitatively, the distinction between small and
large-$J$ is determined by whether one or both reflecting boundaries
are sensed by the near-critical symmetry-broken state.  The large-$J$
transition resembles Bose-Einstein condensation in the sense that
either $n_0$ or $n_J$ accounts for a finite fraction of the particles,
with the remainder ``thermalized'' with an exponential distribution in
$j$ into the interior, at ``temperature'' self-consistently determined
by $q/N = 1/ g$.  To understand the nature of these transitions beyond
mean-field theory we introduce the operator and path-integral
representations of the master equation and its solutions.

\section{Operator, state, and path integral representations}
\label{sec:Op_State_PI}

\subsection{Operators, states, and time evolution} 

The operator representation of master equations from
reaction-diffusion theory~\cite{Mattis:RDQFT:98,Cardy:FTNEqSM:99}
begins by introducing raising and lowering operators
${a}_j^{\dagger}$, $a_j$ for each site on the lattice, with the
commutation relations of orthogonal quantum harmonic oscillators
$\left[ a_j, {a}_{j^{'}}^{\dagger} \right] \equiv {\delta}_{j,j^{'}}
$.  These define a Hilbert space through their action on a ``vacuum''
state ${a}_j \left| 0 \right) \equiv 0 , \forall j$, and its conjugate
$\left( 0 \right| {a}_j^{\dagger} \equiv 0 , \forall j$.

Number states indexed by the vector $n$ are defined through the action
of the raising operators 
\begin{equation}
  \left| n \right) \equiv
  \prod_{j = 0}^{J}
  {
    \left( 
      {a}_j^{\dagger}
    \right)
  }^{n_j}
  \left| 0 \right) ,
\label{eq:number_states_def}
\end{equation}
and differ from quantum-mechanical number states in being normalized
with respect to a universal Glauber state
\cite{Mattis:RDQFT:98,Cardy:FTNEqSM:99}
\begin{equation}
  \left( 0 \right|
  e^{\sum_j a_j}
  \left| n \right) = 1 , 
  \forall n . 
\label{eq:glauber_norm_def}
\end{equation}
The number operator for each $j$ is defined as ${\hat{n}}_j \equiv
{a}_j^{\dagger} {a}_j$, and extracts the appropriate coefficient from
$n$ in the Glauber norm, 
\begin{equation}
  \left( 
    0 
  \right| 
    \exp \left( \sum_{j'} a_{j'} \right) 
    {\hat{n}}_j
  \left|
    n 
  \right) = n_j . 
\label{eq:num_op_action}
\end{equation}

A classical distribution $P \! \left( n \right)$ has the state
representation in the $n$ basis 
\begin{equation}
  \left| \psi \right) \equiv
  \sum_{n}
  P \left( n \right)
  \left| n \right) . 
\label{eq:psi_def}
\end{equation}
$\left| \psi \right)$ is equivalent to a generating function $f \!
\left( z \right)$ of a $\left( J+1 \right)$-component complex vector
$z$, under the association ${a}_j^{\dagger} \leftrightarrow z_j$,
${a}_j \leftrightarrow \partial / \partial z_j$.  Glauber
normalization is equivalent to a prescription for shifting $z_j
\rightarrow z_j + 1$, and evaluating the resulting function at $z_j =
0 , \forall j$.  A thorough treatment of these methods for handling
generating functions and functionals is provided in
Ref.~\cite{Smith:evo_games:11} in the context of the 
analysis of master equations for evolutionary games. However for 
the treatment that follows below, we need only the definitions provided 
above in order to proceed.

The master equation~(\ref{eq:master_eqn}) corresponds to state
evolution equation known as the {\em Liouville equation}
\begin{equation}
  \frac{
    d 
  }{
    dt
  }
  \left| \psi \right) = 
  - \Omega
  \left| \psi \right) , 
\label{eq:psi_evolve_eqn}
\end{equation}
in which the nonlinear, diffusive {\em Liouville operator} that
evolves the state in time is given by 
\begin{equation}
  \Omega = 
  q 
  \sum_{j=0}^{J-1}
  \left(
    {a}_{j+1}^{\dagger} - 
    {a}_j^{\dagger}
  \right)
  \left[
    \left(
      1 + 
      \frac{
        {\hat{n}}_0
      }{
        q
      }
    \right)
    {a}_{j+1} - 
    \left(
      1 + 
      \frac{
        {\hat{n}}_J
      }{
        q
      }
    \right)
    {a}_j
  \right]
\label{eq:Omega_form}
\end{equation}
Here $I= P \equiv q $ as mentioned earlier. The differential 
equation~(\ref{eq:psi_evolve_eqn}) 
is formally
reduced to quadrature to give the time-dependent state relation
\begin{equation}
  \left| 
    {\psi}_t 
  \right) \equiv 
  e^{- \Omega t}
  \left| 
    {\psi}_0  
  \right) . 
\label{eq:time_evolved_psi}
\end{equation}
Normalization of $P \! \left( n \right)$ and the number states $\left|
n \right)$ implies $ \left( 0 \right| \exp \left( \sum_j a_j \right)
\left| {\psi}_t \right) = 1 , \forall t $.  We further recognize the
exogenous catalytic strength $q \equiv 1 / \tau$ as defining a natural
timescale, and the natural coupling $g \equiv N / q$ in
Eq.~(\ref{eq:Omega_form}), as before.

\subsection{Coherent-state expansion and path integral}

At weak nonlinearity (small $g$), it is both intuitive and computationally
efficient to expand solutions to Eq.~(\ref{eq:time_evolved_psi}) in
eigenvectors of the annihilation operators
$a_j$~\cite{Cardy:FTNEqSM:99}, which are the Poisson distributions in
$n_j$.  We start with a normalized initial state arbitrarily
parametrized by mean occupation numbers
\begin{equation}
  \left| {\psi}_0 \right) = 
  \exp 
  \left( 
    \sum_j 
    {\bar{n}}_j 
    \left( 
      a_j^{\dagger} - 1 
    \right)
  \right)
  \left| 0 \right) , 
\label{eq:psi_init_cond}
\end{equation}
in which judicious choice of the ${\bar{n}}_j$ cancels surface terms
associated with transients.  (Self-consistency of these parameters
with stationarity under $\Omega$ may be used from the operator
representation to obtain moments of $P \! \left( n \right)$, as was
done in Ref.~\cite{Sasai:gene_exp:03}, though we will proceed directly
to the time-dependent field action here.)  To form a basis for
coherent-state expansion (again, see \cite{Cardy:FTNEqSM:99} for
detials of this procedure) at increments of time, we introduce a
complex-valued vector field ${\phi}_t \equiv {\left( {\phi}_0 ,
{\phi}_1, \ldots, {\phi}_J \right)}_t^T$, and its adjoint
${\phi}^{\dagger}_t$.  At a set of $t^{'} = k \Delta t$, we insert the
representation of identity
\begin{equation}
  \int 
  \frac{
    d{\phi}^{\dagger}_{t^{'}} \, d{\phi}_{t^{'}}
  }{
    \pi
  }
  e^{- {\phi}^{\dagger}_{t^{'}} \cdot {\phi}_{t^{'}}}
  e^{a^{\dagger} \cdot {\phi}_{t^{'}}}
  \left| 0 \right)
  \left( 0 \right|
    e^{{\phi}^{\dagger}_{t^{'}} \cdot a} = 
  \sum_{n}
  \left| n \right)
  \left( n \right| = 
  I 
\label{eq:one_for_insertion}
\end{equation}
into $ \left( 0 \right| \exp \left( \sum_j a_j \right) \left| {\psi}_t
\right) $, expressed through Eq.~(\ref{eq:time_evolved_psi}) and
Eq.~(\ref{eq:psi_init_cond}) as $ \left( 0 \right| \exp \left( \sum_j
a_j \right) e^{- \Omega t} \exp \left( \sum_j {\bar{n}}_j \left(
a_j^{\dagger} - 1 \right) \right) \left| 0 \right) $.  Though the
coherent states are overcomplete, phase cancellations in
Eq.~(\ref{eq:one_for_insertion}) leave the proper complete number
basis at each $t^{'}$.

By now-standard procedures~\cite{Mattis:RDQFT:98,Cardy:FTNEqSM:99} we
recognize that the fields $\phi$ and ${\phi}^{\dagger}$ have somewhat
different roles, with fluctuations in $\phi$ about its mean value
corresponding roughly to fluctuations in number, and those in
${\phi}^{\dagger}$ sampling moments of the generating functional
$\left| \psi \right)$.  Thus we expand the complex-conjugate
coefficients ${\phi}^{*}_j$ of the (row) vector ${\phi}^{\dagger}$ at
each time as ${\phi}^{*}_j \equiv {\tilde{\phi}}_j + 1$ at each $t$, 
with shorthand
${\phi}^{\dagger}_t \equiv {\tilde{\phi}}_t + 1$, leaving $\phi$ to be
determined physically.  The resulting normalized generating functional
has the path-integral representation

\begin{equation}
  \left( 0 \right| 
  \exp \left( \sum_j a_j \right) 
  \left| {\psi}_t \right) = 
  \int
  {\cal D} {\tilde{\phi}} \, 
  {\cal D} \phi
  e^{- \int dt L}
  e^{
    {\tilde{\phi}}_0 
    \cdot 
    \left( 
      \bar{n} - {\phi}_0
    \right)
  } , 
\label{eq:funct_int_rep}
\end{equation}
in which the diffusive ``Lagrangian'' is 
\begin{equation}
  L 
  \left(
    {\tilde{\phi}} , \phi
  \right) = 
  {\tilde{\phi}} \cdot
  \frac{
    \partial \phi
  }{
    \partial t
  } + 
  \Omega 
  \left( 
    {\tilde{\phi}} + 1 , \phi 
  \right) . 
\label{eq:lagrangian_form}
\end{equation}

The Liouville operator~(\ref{eq:Omega_form}) has induced a
complex-valued function of fields $\Omega \left( {\tilde{\phi}} + 1 ,
\phi \right)$ through the substitution ${a}_j^{\dagger} \rightarrow
{\tilde{\phi}}_j + 1$, $a_j \rightarrow {\phi}_j$.  We will see that,
up to care with signs and contours of integration that depend on what
we wish to extract from this function, it behaves as the equivalent of
a Hamiltonian for an equilibrium system, with a few structural
differences characteristic of stochastic processes(elaborated also in
~\cite{Smith:evo_games:11}).  The measure
${\cal D} {\tilde{\phi}} \, {\cal D} \phi$ in
Eq.~(\ref{eq:funct_int_rep}) is defined formally by the
skeletonization procedure for insertion of the coherent states, but in
practice is usually defined implicitly by perturbation theory in the
diffusive Green's function\footnote{See Ref.~\cite{Kamenev:DP:01} for
a thorough development of fluctuations in both Doi-Peliti formulation
of stochastic processes, and two-field methods more generally
including the Keldysh~\cite{Keldysh::65} and Martin-Siggia-Rose
methods~\cite{Martin:MSR:73}.  In these theories, Green's functions
describe the response of either the observable or moment-sampling
fields ($\phi$ or ${\phi}^{\dagger}$) to point sources.  They are the
basis for Langevin and other expansions for the treatment of noise.}.

Linearization of $\Omega$ in either Eq.~(\ref{eq:psi_evolve_eqn}) or
Eq.~(\ref{eq:funct_int_rep}) ({\it i.e.} keeping only terms linear in ${\tilde \phi}$) provides the natural expansion in
independent collective fluctuations of the master equation and gives results for expectation values which are identical with the mean-field results presented earlier.  In the
field form~(\ref{eq:lagrangian_form}) it further provides a convenient
and intuitive background-field expansion, in which the backgrounds
represent locally best-fit Poisson distributions with mean number
equal to ${\phi}^{\ast}_j {\phi}_j$ for each component $j$.

\subsection{Structure of reaction-diffusion Lagrangians}

To make use of the form of $\Omega$, we introduce two projection
matrices onto the catalytic sites $j=0,J$, $P_{\pm}$
with components   
$ {\left( P_{+} \right) }_{jj^{'}} \equiv 
  {\delta}_{j,J} {\delta}_{j^{'},J}$, and 
$ {\left( P_{-} \right) }_{jj^{'}} \equiv 
  {\delta}_{j,0} {\delta}_{j^{'},0}$, and 
linear-diffusion matrices
\begin{equation}
  D_{+} \equiv 
  \left[
    \begin{array}{rrrrr}
      1  & \mbox{ } &        &    &   \\
      -1 &          & \ddots &    &   \\
         &          & \ddots &  1 &   \\
         &          &        & -1 & 0   
    \end{array}
  \right] , 
\label{eq:plus_diff_op_def}
\end{equation}
\begin{equation}
  D_{-} \equiv 
  \left[
    \begin{array}{rrrrr}
      0 & -1 & \mbox{ } &        &     \\
        &  1 &          & \ddots &     \\
        &    &          & \ddots & -1  \\
        &    &          &        & 1   
    \end{array}
  \right] , 
\label{eq:minus_diff_op_def}
\end{equation}
corresponding to phosphorylation and dephosphorylation transitions,
respectively.  

Noting that for $1^T$ the row vector of all ones, $1^T D_{\pm} \equiv
0$, we can use ${\phi}^{\dagger}$ or ${\tilde{\phi}}$ as it is
convenient, to write 
\begin{eqnarray}
  \frac{1}{q}
  \Omega 
  \left( 
    {\phi}^{\dagger} , \phi 
  \right) 
& = & 
  \left(
    1 + \frac{1}{q}
    {\phi}^{\dagger} P_{-} \phi
  \right)
  {\phi}^{\dagger} D_{-} \phi 
\nonumber \\
& & 
  \mbox{} + 
  \left(
    1 + \frac{1}{q}
    {\phi}^{\dagger} P_{+} \phi
  \right)
  {\phi}^{\dagger} D_{+} \phi . 
\label{eq:Omega_matrix_form}
\end{eqnarray}

We extract the overall $N$-dependence of the action in the path
integral by descaling time with the definition $dz \equiv dt q \equiv
dt / \tau$, and rescaling the Lagrangian to a Lagrangian density per
particle, to write
\begin{equation}
  e^{- \int dt L} = 
  e^{- N \int dz \hat{L}} , 
\label{eq:descale_time}
\end{equation}
with $\hat{L} \equiv L / qN$.  If we similarly descale the field $\phi
\rightarrow \hat{\phi} \equiv \phi / N$, and the Liouvillian $\Omega
\rightarrow \hat{\Omega} \equiv \Omega / qN$, we have the Lagrangian
density in terms of the natural coupling $g=N/q$:
\begin{equation}
  \hat{L} = 
  {\tilde{\phi}} \, 
  {\partial}_z \hat{\phi} + 
  \hat{\Omega} 
  \left( 
    {\phi}^{\dagger} , \hat{\phi} 
  \right) , 
\label{eq:spec_nat_Lagrangian}
\end{equation}
where 
\begin{eqnarray}
  \hat{\Omega} 
  \left( 
    {\phi}^{\dagger} , \hat{\phi} 
  \right) 
& = & 
  \left(
    1 + g
    {\phi}^{\dagger} P_{-} \hat{\phi}
  \right)
  {\phi}^{\dagger} D_{-} \hat{\phi}
\nonumber \\
& & 
  \mbox{} + 
  \left(
    1 + g
    {\phi}^{\dagger} P_{+} \hat{\phi}
  \right)
  {\phi}^{\dagger} D_{+} \hat{\phi} . 
\label{eq:spec_nat_potential}
\end{eqnarray}
Note that the natural fields define the relative number operator
${\phi}^{\dagger}_j {\hat{\phi}}_j = n_j / N \equiv {\nu}_j$,
satisfying $\left< \sum_{j=0}^J {\nu}_j \right> \equiv 1$.  Now not
only are the fields ${\phi}^{\dagger}$ and $\hat{\phi}$ expanded about
different backgrounds, comparable fluctuations of $\tilde{\phi}$ and
$\hat{\phi}$ correspond to fluctuations of ${\phi}^{\dagger}$ and
$\phi$ on scales differing by $N$, with large $N$ defining the domain
of perturbation theory.

To expand the functional integral~(\ref{eq:funct_int_rep}) in Gaussian
fluctuations, we further separate out mean values from the fields,
introducing notation 
$\tilde{\phi} \equiv
\bar{\tilde{\phi}} + \varphi$ (so putting $\bar{{\phi}^{\dagger}} =
\bar{\tilde{\phi}} + 1^T$ and ${\phi}^{\dagger} =
\bar{{\phi}^{\dagger}} + \varphi$).  Using a compact notation
${\hat{\Omega}}^{i}_{j}$ for the tensor of $i$
${\phi}^{\dagger}$-derivatives and $j$ $\hat{\phi}$-derivatives of
${\hat{\Omega}}$, the second-order Taylor expansion in $\varphi$ is
exact: 
\begin{eqnarray}
  \hat{L} 
& = & 
  \bar{\tilde{\phi}} \cdot
  {\partial}_z \hat{\phi} + 
  \hat{\Omega} 
  \left( 
    \bar{{\phi}^{\dagger}} , \hat{\phi} 
  \right) 
\nonumber \\
& & 
  \mbox{} + 
  \varphi \cdot
  \left[
    {\partial}_z \hat{\phi} + 
    {\hat{\Omega}}^{1} 
    \left( 
      \bar{{\phi}^{\dagger}} , \hat{\phi} 
    \right) 
  \right] + 
  \frac{1}{2}
  {\varphi}^2 : 
  {\hat{\Omega}}^{2} 
  \left( 
    \hat{\phi} 
  \right) , 
\nonumber \\
& & 
\label{eq:L_varphi_expand}
\end{eqnarray}
and ${\hat{\Omega}}^{2}$ is independent of $\bar{{\phi}^{\dagger}}$. 

The background $\bar{\tilde{\phi}} \equiv 0$ makes the first line of
Eq.~(\ref{eq:L_varphi_expand}) vanish for general $\hat{\phi}$, and for
more general $\bar{\tilde{\phi}}$ we can expand $\hat{\phi}$ in a
classical background and perturbations, in which the linear order
vanishes at that $\bar{\tilde{\phi}}$.
The $\varphi$-linear term in the second line of
Eq.~(\ref{eq:L_varphi_expand}) enforces a $\delta$-functional if
$\varphi$ is rotated to an imaginary integration contour, and {\em
  negative} eigenvalues of ${\hat{\Omega}}^{2} \left( \hat{\phi}
\right)$ only soften the $\delta$-functional for their corresponding
$\varphi$ eigenvectors with a convergent Gaussian envelope.  We handle
these eigenvalues in perturbation theory with a Hubbard-Stratonovich
transformation~\cite{Weinberg:QTF_II:96} and a Langevin (auxiliary)
field ~\cite{Cardy:FTNEqSM:99}. We see below that in phases 
with no symmetry breaking, the
eigenvalues of ${\hat{\Omega}}^{2} \left( \hat{\phi} \right)$ are all
zero or negative.\footnote{Negative eigenvalues of this Hessian matrix
correspond to decaying modes in the usual sense.  The apparent
divergence caused by the negative sign with which the Liouville
operator $L$ appears is canceled when the complex conjugate fields
${\phi}^{\ast}_j$ -- considered as independent variables of integration
from ${\phi}_j$ -- are rotated to an imaginary integration contour.}

{\em Positive} eigenvalues of ${\hat{\Omega}}^{2} \left( \hat{\phi}
\right)$, of which one appears in the phase of symmetry breaking in
this problem, require different treatment.  They produce a divergent
envelope for the $\delta$-functional integral if $\varphi$ is
integrated along an imaginary contour, while a real contour for a
$\varphi$ eigenvector does not enforce the expected
$\delta$-functional for the corresponding component of the diffusion
equation.  We expect, from experience with Euclidean field theories
for reversible systems, that these eigenvectors signal the existence
of a continuous class of ``approximate'' stationary points generally
termed {\em instantons}~\cite{Coleman:AoS:85}.  $\varphi$ diverges
initially along a real contour, but for the appropriate joint
background of $\hat{\phi}$, nonlinearities in the equations of motion
extend the divergence into a bounded trajectory of locally least $\int
dz \hat{L}$, representing domain flips (a fluctuation that takes the system from one of the bistable phases to the other) in the symmetry-broken phase.  The
integration over the unstable fluctuations of $\varphi$ are not
handled in Gaussian perturbation theory about the static background,
but replaced (with a proper measure term) with the integral over all
time-translates of the approximate stationary solutions.

\subsection{Symmetries and conservations}
\label{subsec:symm_cons}

Foregoing formal treatment of the convergence of Langevin perturbation
theory and its regulation by approximate stationary
points~\cite{Cardy:Instantons:78}, we observe two important global
symmetries of the theory which hold as field identities and also
order-by-order in a large-$N$ expansion.  These are useful in
numerically solving for approximate stationary points in
low-dimensional examples.

The classical equations of motion following from
Eq.~(\ref{eq:L_varphi_expand}) and its equivalent expansion for
$\hat{\phi} = \bar{\hat{\phi}} + {\hat{\phi}}^{'}$ are
\begin{equation}
  {\partial}_z \bar{\hat{\phi}} + 
  {\hat{\Omega}}^{1} 
  \left( 
    \bar{{\phi}^{\dagger}} , \bar{\hat{\phi}} 
  \right) = 
  0 , 
\label{eq:phi_bar_EOM}
\end{equation}
\begin{equation}
  - {\partial}_z \bar{{\phi}^{\dagger}} + 
  {\hat{\Omega}}_{1} 
  \left( 
    \bar{{\phi}^{\dagger}} , \bar{\hat{\phi}} 
  \right) = 
  0 . 
\label{eq:psi_bar_EOM}
\end{equation}
Both are ${\cal O} \! \left( N^0 \right)$, as $\hat{L}$ is defined in
terms only of $z$, $g$, and descaled fields.  The equivalent equations
in terms of ${\phi}^{\dagger}$ and $\phi$, resulting from shifts of
the fields in the measure, generate Ward identities of the theory to
all orders in $N$.  

The transformation ${\phi}^{\dagger} \rightarrow e^{\Lambda}
{\phi}^{\dagger}$, $\hat{\phi} \rightarrow e^{-\Lambda} \hat{\phi}$ at
constant $\Lambda$ is a symmetry of $\hat{L}$ at general
${\phi}^{\dagger}, \hat{\phi}$, whose associated Noether charge is
number: ${\partial}_z \left( {\phi}^{\dagger} \cdot \hat{\phi} \right)
= 0$ as a field equation.  Time-translation is also a symmetry of
$\hat{L}$ whose Noether charge is the potential: ${\partial}_z \left(
\hat{\Omega} \right) = 0$.  Both of these follow immediately as
properties of the classical solutions of Eq's.~(\ref{eq:phi_bar_EOM})
and~(\ref{eq:psi_bar_EOM}).  About backgrounds that are, or converge
to, $\bar{\tilde{\phi}} \equiv 0$, the constraints
$\bar{{\phi}^{\dagger}} \cdot \bar{\hat{\phi}} = 1$ and
${\hat{\Omega}} \left( \bar{{\phi}^{\dagger}} , \bar{\hat{\phi}}
\right) = 0$ specify a $2J$-dimensional subspace of field
configurations in which all classical trajectories must lie.

We further note that, due to the quartic
form~(\ref{eq:spec_nat_potential}), 
\begin{eqnarray}
  \bar{\tilde{\phi}} \cdot 
  {\partial}_z \bar{\hat{\phi}} 
& = & 
  - \bar{\tilde{\phi}} \cdot
  {\hat{\Omega}}^{1} 
  \left( 
    \bar{{\phi}^{\dagger}} , \bar{\hat{\phi}} 
  \right) 
\nonumber \\ 
& = & 
  - {\hat{\Omega}}
  \left( 
    \bar{{\phi}^{\dagger}} , \bar{\hat{\phi}} 
  \right) - 
  \frac{1}{2}
  {\bar{\tilde{\phi}}}^2 : 
  {\hat{\Omega}}^{2} 
  \left( 
    \bar{\hat{\phi}} 
  \right) . 
\label{eq:class_der_reduce}
\end{eqnarray}
The classical action over any stationary trajectory is then 
\begin{eqnarray}
  \hat{L}
  \left( 
    \bar{{\phi}^{\dagger}} , \bar{\hat{\phi}} 
  \right) 
& = & 
  - \frac{1}{2}
  {\bar{\tilde{\phi}}}^2 : 
  {\hat{\Omega}}^{2} 
  \left( 
    \bar{\hat{\phi}} 
  \right) 
\nonumber \\
& = & 
  - g
  \left( 
    \bar{\tilde{\phi}} P_{-} \bar{\hat{\phi}}
  \right)
  \left(
    \bar{\tilde{\phi}} D_{-} \bar{\hat{\phi}}  
  \right) - g 
  \left(
    \bar{\tilde{\phi}} P_{+} \bar{\hat{\phi}}
  \right)
  \left(
    \bar{\tilde{\phi}} D_{+} \bar{\hat{\phi}} . 
  \right) . 
\nonumber \\
& & 
\label{eq:classical_L_simplify}
\end{eqnarray}

The positive eigenvalues of ${\hat{\Omega}}^{2} \left(
\bar{\hat{\phi}} \right)$, which create divergent $\varphi$
fluctuations if we include them in the expansion of
Eq.~(\ref{eq:L_varphi_expand}), correspond to trajectors that take
$\hat{L}$ below the value ($\hat{L} = 0$) of all classical (true)
stationary points.  However, we will see that the nonclassical
``approximate'' stationary points of Eq's.~(\ref{eq:phi_bar_EOM})
and~(\ref{eq:psi_bar_EOM}) produce strictly positive $\hat{L}$, so
that domain flips are suppressed relative to the persistence amplitude
within domains in the symmetry-broken phase.  This will be more transparent
with the representation in terms of action-angle variables introduced
in Sec.~\ref{sec:large_dev}.

\subsection{The background-field expansion to recover mean-field theory}

The relation~(\ref{eq:funct_int_rep}) between the state and path
integral representations of the master equation gives the expected
first moment of the field
\begin{equation}
  \left< 
    {{\phi}_j}_t
  \right> = 
  \left( 0 \right| 
  \exp \left( \sum_j a_j \right) 
  {\hat{n}}_j
  \left| {\psi}_t \right) = 
  \left< {n_j}_t \right> , 
\label{eq:ns_phis_expect}
\end{equation}
where the second angle bracket in Eq.~(\ref{eq:ns_phis_expect})
denotes expectation in the probability $P_t \! \left( n \right)$.
While not a field equation (remember that ${\phi}^{\ast}_j {\phi}_j$
is the combination extracted by ${\hat{n}}_j$), this relates the
classical mean-field solutions 
to the stationary points of $\hat{L}$.  The classical solutions
correspond to the subset of stationary
points~\cite{Eyink:action:96,Mattis:RDQFT:98}
\begin{equation}
  {
    \left.
      \frac{
        \partial L
      }{
        \partial {\tilde{\phi}}
      }
    \right|
  }_{
    {\tilde{\phi}} \equiv 0
  } = 0
\label{eq:MFT_def_eqn}
\end{equation}
which solve Eq.~(\ref{eq:phi_bar_EOM}) at $\bar{{\phi}^{\dagger}}
\equiv 1$.  These need not be time-independent, and include the full
suite of classical diffusion trajectories.  However, if the
${\bar{n}}_j$ in the initial condition~(\ref{eq:psi_init_cond}) are
set to the steady-state values, they satisfy detailed balance under
the ratio of catalytic rates corresponding to
Eq.~(\ref{eq:x_def_gen_symtop}) (at $\lambda = 0$ in this case):
\begin{equation}
  x \equiv 
  e^{\xi} = 
  \frac{
    1 + g {\bar{\hat{\phi}}}_J 
  }{
    1 + g {\bar{\hat{\phi}}}_0 
  } . 
\label{eq:x_def}
\end{equation}
The fields themselves satisfy 
\begin{equation}
  {\bar{\phi}}_j = 
  x^j
  {\bar{\phi}}_0 , \;  
  0 \le j \le J , 
\label{eq:0_J_x_reln_bf}
\end{equation}
per Eq.~(\ref{eq:0_J_x_reln}), and the remaining solutions for the
coupling follow.  

\section{Fluctuations about static mean fields}
\label{sec:flucts}

\begin{figure}[ht]
  \begin{center} 
  \includegraphics[scale=0.325]{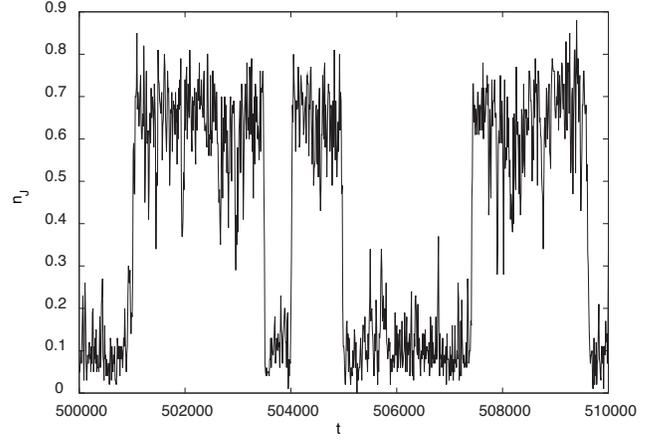}
  \caption{
  Simulation results for a population history, represented by $n_J$
  (number of full-phosphorylated particles). We have used a system of
  $100$ particles and a $g$-value of $4.08$.
    \label{fig:wells_timeseries} 
  }
  \end{center}
\end{figure}

Fig.~\ref{fig:wells_timeseries} shows the general character of
timeseries for the population as represented by the number $n_J$, in a
phase with relatively strong symmetry breaking ($g=4.08$ for $J=2$. As
a reference the phase transition occurs at $g=3$).  A timeseries is
characterized by dense fluctuations about the mean value, in which
$n_J$ remains near the mean-field value, punctuated by occasional
large excursions that shift the mean.  In this section we will
consider the Gaussian-order approximation to the dense fluctuations
about the mean.  In Sec.~\ref{sec:large_dev} we return to
qualitatively different methods to handle the rare events which change
mean population state.

\subsection{Diffusion eigenvalues and eigenvectors}

We compute the noise spectrum by further expanding
Eq.~(\ref{eq:L_varphi_expand}) about $\bar{\tilde{\phi}} \equiv 0$,
defining $\hat{\phi} \equiv \bar{\hat{\phi}} + {\hat{\phi}}^{'}$ and
letting $\bar{\hat{\phi}}$ be a constant solution to
Eq.~(\ref{eq:MFT_def_eqn}), so that the linear term in $\varphi$
vanishes.  Using $\hat{L} \left( 0, \bar{\hat{\phi}} \right) = 0$, the
second-order expansion defines the Gaussian kernel, with the form
\begin{equation}
  \hat{L} = 
  \varphi 
  D_0 
  {\hat{\phi}}^{'} + 
  \varphi
  D_2
  {\varphi}^T + 
  \mbox{h.o.} 
\label{eq:MFT_2nd_ord_exp}
\end{equation}
and higher-order terms ($\mbox{h.o.} $) are left for perturbative
expansion.  The diffusion kernel governing ${\hat{\phi}}^{'}$ in
Eq.~(\ref{eq:MFT_2nd_ord_exp}) is
\begin{eqnarray}
  D_0 
& = & 
  {\partial}_z + 
  \left( 
    1 + g {\bar{\hat{\phi}}}_0
  \right)
  D_{-} + 
  \left( 
    1 + g {\bar{\hat{\phi}}}_J
  \right)
  D_{+} 
\nonumber \\
& & 
\mbox{} + 
  g 
  \left(
    D_{-} 
    \bar{\hat{\phi}}
    1^T 
    P_{-} + 
    D_{+} 
    \bar{\hat{\phi}}
    1^T 
    P_{+}
  \right) , 
\label{eq:bare_diff_prop_MFT}
\end{eqnarray}
while the kernel controlling the constraint field $\varphi$ is 
\begin{equation}
  D_2 = 
  \frac{g}{2}
  \left\{
    \left(
      D_{-} 
      \bar{\hat{\phi}}
      {\bar{\hat{\phi}}}^T 
      P_{-} + 
      D_{+} 
      \bar{\hat{\phi}}
      {\bar{\hat{\phi}}}^T 
      P_{+}
    \right) + 
    \mbox{transpose}
  \right\} . 
\label{eq:bare_noise_kernel_MFT}
\end{equation}

Remarkably, about general normalized solutions to
Eq.~(\ref{eq:0_J_x_reln}), the kernel~(\ref{eq:bare_noise_kernel_MFT})
has only two nonzero eigenvalues ${\lambda}_{\pm}$, with eigenvectors
$v_{\pm}$: 
\begin{equation}
  D_2 
  v_{\pm} = 
  \frac{g}{2}
  {\lambda}_{\pm}
  v_{\pm} . 
\label{eq:D_2_nonzero_eigs}
\end{equation}
We construct $v_{\pm}$ from convenient, orthonormal ``center'' and
``edge'' components,
\begin{equation}
  {
    \left( v_c \right)
  }_j = 
  \sqrt{
    \frac{
      \sinh \left( \xi \right)
    }{
      \sinh \left[ \left( J-1 \right) \xi \right]
    }
  }
  \left( 
    x^{j - J/2} 
  \right) , 
  1 \le j \le J-1 , 
\label{eq:v_c_comps}
\end{equation}
and zero otherwise, and 
\begin{equation}
  {
    \left( v_e \right)
  }_{j = \left( J/2 \pm J/2 \right)} = 
  \frac{
    \pm 
    x^{
      \pm 
      \left( \frac{J-1}{2} \xi \right)
    }
  }{
    \sqrt{
      2 \cosh \left[ \left( J-1 \right) \xi \right]
    }
  }
\label{eq:v_e_comps}
\end{equation}
and zero otherwise.

A term that appears in the solution for the eigenvalues is abbreviated 
\begin{equation}
  {\cal R}
  \left( \xi \right) \equiv 
  \sqrt{
    1 + 
    \tanh \left[ \left( J - 1 \right) \xi \right]
    \tanh \left( \frac{1}{2} \xi \right) 
  } , 
\label{eq:rad_tanh_notation}
\end{equation}
in terms of which 
\begin{equation}
  {\lambda}_{\pm} = 
  2
  \left\{
    \pm
    {\cal R}
    \left( \xi \right) - 
    1
  \right\}
  \cosh \left[ \left( J - 1 \right) \xi \right]
  {
    \left(
      \frac{
        \sinh \left( \frac{1}{2} \xi \right)
      }{
        \sinh \left( \frac{J+1}{2} \xi \right) 
      }
    \right)
  }^2 , 
\nonumber \\
\label{eq:lambdas_noise_broken}
\end{equation}
and the orthonormal eigenvectors
\begin{equation}
  v_{\pm} = 
  \frac{
    1
  }{
    \sqrt{
      2 {\cal R} \left( \xi \right)
    }
  }
  \left\{
    v_c 
    \sqrt{{\cal R} \left( \xi \right) \pm 1} \mp
    v_e
    \sqrt{{\cal R} \left( \xi \right) \mp 1}
  \right\} . 
\label{eq:v_pm_solve}
\end{equation}

For $g \le g_c$, only $\xi = 0$ is consistent, and we get
${\lambda}_{+} \equiv 0$, 
\begin{equation}
  {\lambda}_{-} = 
  -
  {
    \left(
      \frac{ 2 }{ J+1 }
    \right)
  }^2 , 
\label{eq:lambda_noise_sym}
\end{equation}
with eigenvector $v_{-} = v_e$. 
This algebraic result emphasizes the efficiency of expanding about
Poisson backgrounds for weakly perturbed stochastic processes.  The
only deviation from Poisson which must be handled perturbatively comes
from a single mode of $\varphi$, whose fluctuations represent
exchanges between $n_0$ and $n_j$ by Eq.~(\ref{eq:v_e_comps}).  These
are of course the noise in the catalytic rates that feeds back into
the distribution as a whole.

\subsection{Hubbard-stratonovitch transformation about the symmetric
phase} 

Rather than complete the square in Eq.~(\ref{eq:MFT_2nd_ord_exp})
(\'{a} la Onsager and Machlup~\cite{Onsager:Machlup:53}), which is
cumbersome for one eigenvector, we introduce into
Eq.~(\ref{eq:funct_int_rep}) an auxiliary-field representation of
unity at each time:
\begin{equation}
  1 = 
  {\cal N}
  \int 
  {\cal D} \tilde{\zeta}
  e^{
    - \frac{N}{2g} \int dz 
    {\tilde{\zeta}}^2
  } , 
\label{eq:langevin_unit_insert}
\end{equation}
in which ${\cal N}$ is a time- and field-independent normalization.
Shifting the auxiliary field $\tilde{\zeta}$ (a symmetry of the
measure), we introduce the physical Langevin field $\zeta$ as 
\begin{equation}
  \tilde{\zeta} \equiv 
  \zeta - 
  g
  \sqrt{- {\lambda}_{-}}
  \left( 
    \varphi \cdot
    v_{-}
  \right) .
\label{eq:tilde_zeta_shift}
\end{equation}
The net effect on $\hat{L}$ is the shift
\begin{eqnarray}
  \hat{L} 
& \rightarrow & 
  \frac{1}{2g} {\tilde{\zeta}}^2 + 
  \hat{L} 
\nonumber \\
& \approx & 
  \frac{1}{2g} {\zeta}^2 + 
  \varphi
  \left(
    D_0 
    {\hat{\phi}}^{'} - 
    \sqrt{- {\lambda}_{-}}
    v_{-} \zeta
  \right) + 
  \frac{g}{2}
  {\lambda}_{+}
  {
    \left( 
      \varphi \cdot 
      v_{+}
    \right)
  }^2 . 
\nonumber \\
\label{eq:L_Langevin_shift}
\end{eqnarray}
$\zeta$ is $\delta$-correlated in $z$ with weight $g / N$, 
\begin{equation}
  \left<
    {\zeta}_z 
    {\zeta}_{z^{\prime}} 
  \right> = 
  \frac{g}{N}
  \delta 
  \left( z - z^{\prime} \right) , 
\label{eq:zeta_MFT_fluct}
\end{equation}
and drives the field ${\hat{\phi}}^{'}$ through the inverse of $D_0$,
acting on $v_{-}$.  In the symmetric phase ${\lambda}_{+} = 0$ and
this is all there is to the bare noise kernel; in the symmetry-broken
phase we must still handle (by other means) the term $\varphi \cdot
v_{+}$, which however remains orthogonal to the $v_{-}$ in the
Langevin term.  Integration over $\varphi$ in the symmetric phase
produces
\begin{equation}
  {\hat{\phi}}^{'}_z = 
  \sqrt{- {\lambda}_{-}}
  \int_0^z dz^{\prime}
  G_0 \left( z, z^{\prime} \right) 
  v_{-} 
  {\zeta}_{z^{\prime}} 
\label{eq:MFT_phi_from_zeta}
\end{equation}
as a field equation to Gaussian order, in which $G_0 \left( z, z^{'}
\right)$ is defined in terms of Eq.~(\ref{eq:bare_diff_prop_MFT}) by
\begin{equation}
  D_0 \, 
  G_0 \! \left( z, z^{\prime} \right) \equiv 
  \delta 
  \left( z - z^{\prime} \right) . 
\label{eq:bare_G_0_def}
\end{equation}
Note that from Eq. ~\ref{eq:MFT_phi_from_zeta} we see that
$\left<{\hat{\phi}}^{'}_z\right> = 0$, which implies that there are no 
corrections to the mean-field result for the expectation value in the
symmetric phase.

\subsection{Fluctuations about the symmetric order parameter}

As an example we compute to lowest order the fluctuations in the order
parameter about the symmetric phase, where the diffusive Green's
function is easy to compute in closed form.  Application of the number
operators first to the basis $\left| n \right)$ and then to the
coherent-states in Eq.~(\ref{eq:funct_int_rep}) yields the connected
component of the variance (expressed in descaled fields)
\begin{equation}
  \frac{
    \left<
      {
        \left( 
          n_J - n_0
        \right)
      }^2
    \right> - 
    {
      \left<
        n_J - n_0
      \right>
    }^2
  }{
    2N / 
    \left( J + 1 \right)
  } = 
  1 + 
  N 
  \left( J + 1 \right)
  \left<
    {
      \left(
        \frac{
          {\hat{\phi}}^{'}_J - {\hat{\phi}}^{'}_0
        }{
          \sqrt{2}
        }
      \right)
    }^2_z
  \right> . 
\label{eq:ord_parm_flucts}
\end{equation}
From the field equation~(\ref{eq:MFT_phi_from_zeta}) and the
correlator~(\ref{eq:zeta_MFT_fluct}) we obtain the ${\hat{\phi}}^{'}$
noise 
\begin{widetext}
\begin{equation}
  \left<
    {
      \left(
        \frac{
          {\phi}^{'}_J - {\phi}^{'}_0
        }{
          \sqrt{2}
        }
      \right)
    }^2_z
  \right> = 
  - {\lambda}_{-}
  \frac{g}{N}
  \int_0^z dz^{\prime}
  {
    \left(
      \frac{
        \left[
          \begin{array}{ccccc}
            -1 & 0 & \cdots & 0 & 1
          \end{array}
        \right] 
      }{
        \sqrt{2}
      }
      \cdot
      G_0 \left( z, z^{\prime} \right) \cdot
      v_{-}
    \right)
  }^2  , 
\label{eq:MFT_phi_pr_sqr}
\end{equation}
\end{widetext}
and it remains only to compute the mode expansion of the diffusion
kernel in $D_0$ from Eq.~(\ref{eq:bare_diff_prop_MFT}) in the uniform
background $\bar{\hat{\phi}}=1 / \left( J+1 \right)$.

The symmetric-phase $D_0$ contains a symmetric linear diffusion matrix
with endpoint corrections, so its eigenvectors $v_k$ have components $
{\left( v_k \right)}_j = {\cal N}_k \cos \left[ {\theta}_k +
{\kappa}_k \left( j - J / 2 \right) \right] $, with ${\cal N}_k$ a
normalization.  The eigenvalues are immediate on the interior sites, 
\begin{equation}
  {\lambda}_k =
  4 
  \left( 
    1 +
    \frac{g}{J+1} 
  \right)
  {\sin}^2
  \frac{
    {\kappa}_k
  }{
    2
  } , 
\label{eq:sym_eig_values}
\end{equation}
and consistency of the interior with the endpoint corrections
then determines ${\kappa}_k$ and ${\theta}_k$. 

{\bf Either}
$ \sin \left( J + 1 \right) {\kappa}_k / 2 = 0
  \Rightarrow 
  {\kappa}_k = \pi k / \left( J + 1 \right)
$, $k$ even, and 
$ {\theta}_k = 0
$, {\bf or} 
$ \cos {\theta}_k = 0 \Rightarrow {\theta}_k = \pi / 2$ and
  ${\kappa}_k$ solves the matching equation
\begin{equation}
  \tan 
  \frac{
    {\kappa}_k
  }{
    2
  } = 
  \frac{
    g
  }{
    J + 1 + 2 g 
  }
  \tan \frac{J+1}{2} 
  {\kappa}_k . 
\label{eq:sym_wn_sols}
\end{equation}

Eq.~(\ref{eq:sym_wn_sols}) is regular for $k \ge 3$ odd, and creates
only a small wavenumber shift from the free diffusion solution,
leaving $ {\kappa}_k \approx \pi k / \left( J + 1 \right)$.  The
important mode for critical behavior is $k=1$ as $g \rightarrow g_c$,
where ${\kappa}_1 \rightarrow 0$ as
\begin{equation}
  \frac{
    {\kappa}_1^2
  }{
    6 
  } \rightarrow 
  \frac{
    1 
  }{
    {
      \left( J+1 \right)
    }^2
  }
  \left( 
    \frac{1}{g} - 
    \frac{1}{g_c}
  \right) . 
\label{eq:crit_zero_wn}
\end{equation}

In terms of these the modal expansion of the free Green's function is 
\begin{equation}
  G_0 \left( z, z^{\prime} \right) = 
  \Theta \! 
  \left( z - z^{\prime} \right)
  \sum_{k=1}^{J+1}
  e^{
    - {\lambda}_k 
    \left( z - z^{\prime} \right)
  }  
  v_k
  v_k^T . 
\label{eq:G0_mode_expand}
\end{equation}
Only odd-$k$ modes from $G_0$ couple to $v_{-}$ in
Eq.~(\ref{eq:MFT_phi_pr_sqr}), by symmetry, and the wavenumber sums
from $k \ge 3$ are easily approximated with a two-dimensional $k$
integral.  We note that for all these modes ${\cal N}_k^2 \approx 2 /
\left( J+1 \right)$, and the only values that contribute to the inner
product come from $j = \pm J/2$, giving ${\sin}^2 \! \left( {\kappa}_k
J/2 \right) \approx 1$.  Thus the nonsingular modes in the diffusion
kernel provide a smooth background approximately linear in $g$.  

The leading contribution from the $k=1$ mode occurs when it is present
in both factors of $G_0$.  For small $g - g_c$ this mode is almost
linear in $j$, with normalization ${\cal N}_1^{-2} \rightarrow
{\kappa}_1^2 \sum_{j = -J/2}^{J/2} j^2$.  Evaluating this singular
term separately, with Eq.~(\ref{eq:sym_eig_values}) for the eigenvalue
and Eq.~(\ref{eq:crit_zero_wn}) for the limiting value of the
wavenumber, and then combining with the background from the regular
modes, we obtain the approximation
\begin{widetext}
\begin{equation}
    \frac{
    \left<
      {
        \left( 
          n_J - n_0
        \right)
      }^2
    \right> - 
    {
      \left<
        n_J - n_0
      \right>
    }^2
  }{
    2N / 
    \left( J + 1 \right)
  } \approx
  1 + 
  \frac{8g}{\pi}
  \frac{
    \log 
    \left( J + 1 \right) 
  }{
    \left( J + 1 \right) 
  } + 
  {\cal C}
  \frac{6}{J+1}
  \frac{
    g^2
  }{
    \left( 1 + \frac{g}{J+1} \right)
    \left| g - g_c \right|
  } . 
\label{eq:sym_approx_flucts}
\end{equation}
\end{widetext}
It is convenient to separate the constant 
\begin{equation}
  {\cal C} = 
  \frac{
    {
      \left( J+1 \right)
    }^3
  }{
    J^2
    \left( J-1 \right)
  }
  {
    \left( 
      \frac{12}{J}
      \sum_{j = -J/2}^{J/2}
      {
        \left( 
          \frac{j}{J}
        \right)
      }^2
    \right)
  }^{-2} 
\label{eq:norm_corr_form}
\end{equation}
from the discrete sum for the $k=1$ norm, because ${\cal C}
\rightarrow 1$ at large $J$, but differs somewhat at the smaller $J$
of more likely biological interest.   

\begin{figure}[ht]
  \begin{center} 
  \includegraphics[scale=0.475]{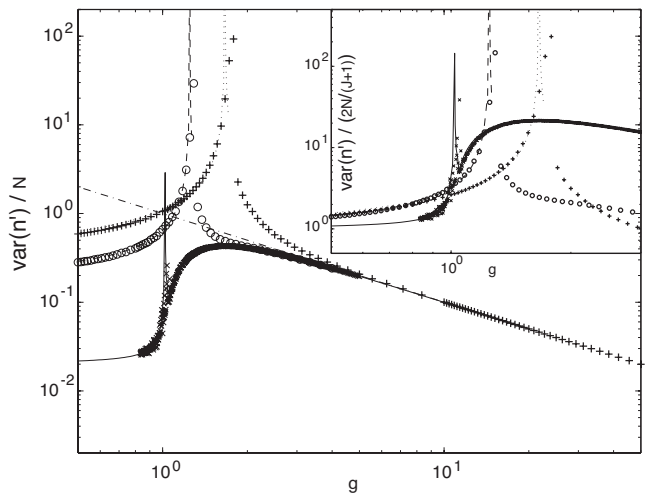}
  \caption{
  Fluctuations in the order parameter scaled for $g$ above and below
  critical.  $\mbox{\em var} \left( n' \right)$ stands for the
  variance $\left<{\left( n_0 - n_J \right)}^2\right> - {\left< n_0 -
  n_J\right>}^2$.  Lines are leading-order expansion of
  Eq.~(\ref{eq:sym_approx_flucts}) in fluctuations about the symmetric
  mean-field solution, continued through $g_c$; symbols from
  simulation, $J+1$ values and markers as in
  Fig.~\ref{fig:MFT_ord_parms_data_paper}.  Large panel shows
  convergence of $\mbox{\em var} \left( n' \right)$ to $N/g$ at all
  $J$.  Inset shows convergence of $\mbox{\em var} \left(
  n' \right)$ to Poisson result $2N / \left( J+1 \right)$ as $g
  \rightarrow 0$.
  \label{fig:MFT_flucts_data_micro} 
  }
  \end{center}
\end{figure}

The approximation~(\ref{eq:sym_approx_flucts}) to the closed-form mode
expansion for the variance of the order parameter is compared to
numerical simulations in Fig.~\ref{fig:MFT_flucts_data_micro}.  We
have continued the analytic expression through the critical point to
show the peak, though the character of the modes rapidly changes as
the distribution becomes skewed in the symmetry-broken phase.  A
similar mode expansion exists in this phase, but requires the
relaxation eigenvectors and eigenvalues for asymmetric diffusion.  We
have not computed these in closed form, and do not pursue them
numerically because in the symmetry-broken phase fluctuations quickly
come to be dominated by the center-of-mass behavior we derive below in
Eq.~(\ref{eq:largeJfluct}).  

Three main observations are important.  First, the singularity in the
variance has the leading-order approximation
\begin{equation}
    \frac{
    \left<
      {
        \left( 
          n_J - n_0
        \right)
      }^2
    \right> - 
    {
      \left<
        n_J - n_0
      \right>
    }^2
  }{
    2N / 
    \left( J + 1 \right)
  } \approx
  \frac{
    \mbox{const} \sim 1
  }{
    \left( J + 1 \right)
    \left| g - g_c \right| 
  } , 
\label{eq:ord_fluct_scaling}
\end{equation}
comparable to that of the mean-field Ising ferromagnet, like the
scaling of the order parameter.  The variance has weight $1 / \left( J
+ 1 \right)$, because the lowest diffusive mode, corresponding to the
average magnetization, is the only collective fluctuation
participating in the phase transition near the critical point.

Second, we see that the weak-coupling scaling of $\left< { \left( n_J
- n_0 \right) }^2 \right> - {\left< n_0 - n_J\right>}^2$ is that of
Poisson noise for an average of $N / \left( J + 1 \right)$ particles
per site.  (This is shown in the inset of
Fig.~\ref{fig:MFT_flucts_data_micro}.)

Third -- and the reason we do not pursue the low-order expansion for
the symmetry-broken-phase two-point function -- we see that for $g \gg
g_c$ the variance $\left< { \left( n_J - n_0 \right) }^2 \right> -
{\left< n_0 - n_J\right>}^2$ goes to a universal form $N/g$ for any
$J$.  The independence of this
scaling regime from $J$ indicates that the particles interact with
one end of the chain and the exogenous catalysis only, suggesting a
strong-coupling limit.

\begin{figure}[ht]
  \begin{center} 
   \includegraphics[scale=0.35]{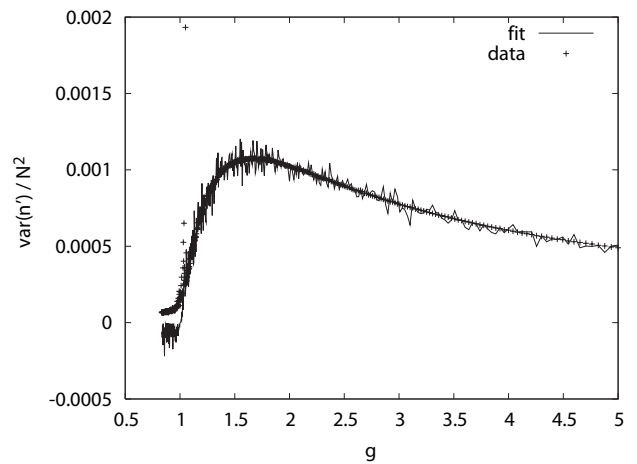}
  \caption{
  The fluctuation in the order parameter for $J+1=100$ and $N=400$ is
  plotted against the fit predicted by Eq. ~\ref{eq:largeJfluct}.
    \label{fig:large_J_fluct} 
  }
  \end{center}
\end{figure}

In addition, for large $J$, the center-of-mass
equation~(\ref{eq:cm}) predicts (for $I= P$), in steady state for the
symmetry-broken phase, that the variance will equal 
\begin{equation}
~ \left<n_e\right> N 
  \left[
    1-\frac{1}{g} -
    \frac{\left<n_e\right>}{N}
  \right]
\label{eq:largeJfluct}
\end{equation}
where $n_e$ signifies whichever of the end sites $0$ or $J$ is
occupied (which depends on which of the two bistable states is
chosen).  In Fig.~\ref{fig:large_J_fluct} we plot the simulation data
for the variance for $J+1=100$ against the estimate from
Eq.~(\ref{eq:largeJfluct}). For the value of $\left<n_e\right>$,  
we simply take the value of the order parameter at the corresponding 
value of $g$. As we see, this explains the form of the
fluctuation spectrum for large J very well. 
The fact that we are able to replace the order parameter 
$\left| \left< n_J - n_0 \right> \right|$ by $n_e$ demonstrates that this
scaling regime is independent of $J$ as well and the particles 
 only interact with one end of the chain all through the 
symmetry-broken phase.

\section{Large excursions}
\label{sec:large_dev}

From the $ 1 / g$ scaling of the Lagrangian term for the Langevin
field in Eq.~(\ref{eq:L_Langevin_shift}) and the presence of unstable
eigenvalues in the fluctuation kernel~(\ref{eq:MFT_2nd_ord_exp}) about
static stationary backgrounds, we anticipate that perturbation theory
containing fluctuations associated with domain flips will not
converge~\cite{Cardy:Instantons:78}, and that the rates for these
large excursions will be computed as an essential singularity with
respect to the perturbation expansion.  The remarkable similarity to
Hamiltonian dynamical systems created by this method for treating
generating functions reduces the problem of estimating both escape
trajectories and first passage times to that of identifiying the
heteroclinic
network~\cite{Gluckheimer:het_cycles:88,Gluckheimer:dyn_sys:02} in the
associated dynamical system.

We construct the expansion appropriate to the second-order
transitions, beginning with the analysis of the exact stationary
points corresponding to solutions of the classical diffusion equation
from arbitrary initial conditions.  From the structure of these
solutions, we identify the ``approximate stationary points''
associated with domain flips, and compute the trajectory and action
for a low-dimensional example numerically.  Comparisons to simulation
suggest that this calculation correctly predicts the leading
exponential dependence on particle number of the residence time in
domains in the symmetry-broken phase.

\subsection{The expansion in semiclassical stationary points}

We define the stationary-point expansion of the path
integral~(\ref{eq:funct_int_rep}) implicitly by the requirement that
the residual perturbation theory converge. In principle we must
include not only the (generally unique) exact stationary point
specified by the initial state $\left| {\psi}_0 \right)$, but also a
sufficient set of ``approximate'' stationary points associated with
states that converge exponentially fast toward $\left| {\psi}_0
\right)$, with respect to prediction of late-time observables.  In
this representation using generating functionals for probability
distributions over spaces of histories, a ``stationary point'' refers
to any full path $\left( \bar{\tilde{\phi}} , \bar{\hat{\phi}}
\right)$ satisfying the classical condition that the linear variations
of $L$ (including time-derivative terms) vanish:
\begin{equation}
    \frac{
      \partial L
    }{
      \partial {\tilde{\phi}}
    } = 0 ; \mbox{ } 
    \frac{
      \partial L
    }{
      \partial \phi
    } = 0 . 
\end{equation}
If $\tilde{\phi} = 0$, the solution of Eq.~(\ref{eq:MFT_def_eqn})
solves the above to recover the mean-field solutions; in this section
we relax the requirement $\tilde{\phi} = 0$ to uncover a larger set of
solutions which result in a nonzero value for $L$.

Formally, then, the removal of the stationary-point contribute leads
to the functional form 
\begin{eqnarray}
\lefteqn{
  \left( 0 \right| 
  \exp \left( \sum_j a_j \right) 
  \left| {\psi}_t \right) =
} & & 
\nonumber \\
& &  
  \sum_{
    \bar{\tilde{\phi}} , \bar{\hat{\phi}}
  }
  e^{
    - N \int \, dz \bar{\hat{L}}
  }
  \int
  {\cal D} \varphi \, 
  {\cal D} {\hat{\phi}}^{'}
  e^{
    - N \int dz 
    \left( \hat{L} - \bar{\hat{L}} \right)
  }
  e^{
    {\tilde{\phi}}_0 
    \cdot 
    \left( 
      \bar{n} - {\phi}_0
    \right)
  } , 
\nonumber \\
& &  
\label{eq:funct_int_rep_sp_exp}
\end{eqnarray}
where $\bar{\hat{L}}$ denotes 
$ \hat{L} \!  
    \left( 
      \bar{\tilde{\phi}} , \bar{\hat{\phi}}
    \right)$. 
As perturbative corrections scale as powers of $1 / N$, by
Eq.~(\ref{eq:MFT_phi_pr_sqr}), to leading order we will treat $\hat{L}
- \bar{\hat{L}}$ as a quadratic form in $\varphi$ and
${\hat{\phi}}^{'}$, of the form~(\ref{eq:MFT_2nd_ord_exp}), now about
more general time-dependent backgrounds.

The formal sum $\sum_{ \bar{\tilde{\phi}} , \bar{\hat{\phi}} }$ is
properly a discrete sum in the number $n \in 0 , \ldots , \infty$ of
domain flips, of a time-ordered integral over their positions $\left\{
z_1 , \ldots , z_n \right\}$.  The integral is
necessary~\cite{Coleman:AoS:85}, because as the domain flips converge
toward true stationary points, the fluctuation generated by time
translation of any solution becomes a null eigenvector of the
functional determinant about that solution.  (This is a variant on
Goldstone's theorem~\cite{Weinberg:QTF_II:96}, associated with the
time-translation symmetry spontaneously hidden by the
instanton~\cite{Smith:evo_games:11}.)  This eigenvector is replaced by
the integral (with a Jacobean), and the remaining functional
determinant is a product of positive eigenvalues, by construction.

We will check that the transition times of the instantons are finite
and that they converge exponentially to the static backgrounds, so
that when they are improbable the dilute-gas sum is well defined.  As
we verify below, the classical solutions all have $\bar{\tilde{\phi}}
\equiv 0$ and zero action, and we denote by $S_0$ the action $N \int
dz \bar{\hat{L}}$ associated with a single instanton.  Letting $1 /
{\zeta}_0$ denote the Jacobean relating the null eigenvalue to the
measure for $z$-translation of the instanton, we recast the sum in
Eq.~(\ref{eq:funct_int_rep_sp_exp}) as
\begin{widetext}
\begin{equation}
  \left( 0 \right| 
  \exp \left( \sum_j a_j \right) 
  \left| {\psi}_t \right) =
  \sum_{n=0}^{\infty}
  \frac{
    e^{
      - n S_0
    }
  }{
    n ! 
  }
  {\bf T} 
  \int \frac{dz_1}{{\zeta}_0} , \ldots , 
  \int \frac{dz_n}{{\zeta}_0}
  \int
  {\cal D} \varphi \, 
  {\cal D} {\hat{\phi}}^{'}
  e^{
    - N \int dz 
    \left( \hat{L} - \bar{\hat{L}} \right)
  }
  e^{
    {\tilde{\phi}}_0 
    \cdot 
    \left( 
      \bar{n} - {\phi}_0
    \right)
  } , 
\label{eq:funct_int_rep_inst}
\end{equation}
\end{widetext}
in which ${\bf T}$ denotes time-ordering in $z$ of the positions of
the instantons.  The presence of $n$ factors of $1 / {\zeta}_0$ in the
$n$-instanton determinant follows from the product structure of
functional determinants and the wide separation of finite supports in
$z$ where the background differs from that of a steady
state~\cite{Coleman:AoS:85}.

Like the computation of the energy shift in the equilibrium
double-well problem, we see that observables relating to persistence
within a domain will receive contributions from the even terms in the
sum over $n$, while those relating to domain flips will receive
contributions from the odd terms.  The likelihood of persistence
decays exponentially in $z$ at early times with rate
\begin{equation}
  r_{\mbox{\scriptsize flip}} = 
  \frac{1}{{\zeta}_0}
  e^{-S_0} . 
\label{eq:flip_rate}
\end{equation}
The computation of the instanton action $S_0$, which is responsible
for the leading exponential dependence of $r_{\mbox{\scriptsize
    flip}}$ is most easily carried out within the complete analysis of
the semiclassical stationary points, beginning with the classical
diffusion solutions.

\subsection{Action-angle variables and the structure of the Hamiltonian}

The equations of motion~(\ref{eq:phi_bar_EOM},\ref{eq:psi_bar_EOM}) in
$\bar{\tilde{\phi}}$ and $\bar{\hat{\phi}}$ do not directly give the
evolution of the physical particle numbers, or efficiently use the
symmetries of Sec.~\ref{subsec:symm_cons}.  To do both, it is
convenient to transform the background fields as
${\bar{{\phi}^{\dagger}}}_j \equiv e^{{\sigma}_j}$, and
${\bar{\hat{\phi}}}_j \equiv {\nu}_j e^{-{\sigma}_j}$.  ${\nu}_j$ is
then the semiclassical approximation to the relative number $\left<
n_j \right> / N$.  This change of variables is equivalent to an
action-angle transformation in classical
mechanics~\cite{Goldstein:ClassMech:01}, and we check in
Ref.~\cite{Smith:evo_games:11} that as well as producing a more
convenient form for the action, it leads to the correct measure for
fluctuations. The 
Lagrangian~(\ref{eq:spec_nat_Lagrangian}) retains a simple kinetic
term, up to a total derivative:
\begin{equation}
  \bar{\hat{L}} = 
  \sigma \cdot
  {\partial}_z \nu + 
  \hat{\Omega} 
  \left( 
    \sigma , \nu
  \right) , 
\label{eq:L_sigma_nu_form}
\end{equation}
and the equations of motion in the new variables become, respectively, 
\begin{equation}
  {\partial}_z
  {\nu}_j = 
  - \frac{
    \partial \hat{\Omega}
  }{
    \partial {\sigma}_j
  } , 
\label{eq:EOM_sigma_nu_form}
\end{equation}
and 
\begin{equation}
  {\partial}_z
  {\sigma}_j = 
  \frac{
    \partial \hat{\Omega}
  }{
    \partial {\nu}_j
  } .  
\label{eq:EOM_nu_sigma_form}
\end{equation}
With the interpretation of the number field $\nu$ as a position, and
$\sigma$ its canonically conjugate momentum, $- \hat{\Omega}$ becomes
the correctly signed Hamiltonian for classical solutions.  The
conservation law $d \hat{\Omega} / dz = 0$ is mathematically a
conservation of energy, but the particular value $\hat{\Omega} \equiv
0$ associated with all stationary points initiated by classical
distributions is a distinctive feature of this stochastic-process
application of Hilbert-space methods.  

The global symmetry whose Noether charge is total number becomes
immediate in action-angle variables.  Defining
\begin{equation}
  \bar{\sigma} \equiv
  \frac{1}{J+1}
  \sum_{j=0}^{J}
  {\sigma}_j , 
\label{eq:subt_ave_sigma}
\end{equation}
the Lagrangian becomes 
\begin{eqnarray}
  \hat{L} 
& = & 
  \bar{\sigma}
  {\partial}_z 
  \left( \sum_{j=0}^{J} {\nu}_j \right) 
\nonumber \\
& & 
  \mbox{} + 
  \sum_{j=0}^{J}
  \left(
    {\sigma}_j - \bar{\sigma}
  \right) 
  {\partial}_z {\nu}_j + 
  \hat{\Omega} 
  \left( 
    \sigma , \nu
  \right) , 
\nonumber \\
& & 
\label{eq:L_sigma_nu_reduced}
\end{eqnarray}
in which $\bar{\sigma}$ multiples the $z$-derivative of conserved
total number in the first line, and only differences ${\sigma}_i -
{\sigma}_j$ appear in either the kinetic term or the $\hat{\Omega}$ of
the second line.

To expose the structure of the associated dynamical system, and to
make the terms in it readily visualizable from the mean-field
diffusive solutions, we perform a final transformation by introducing
the log ratio of particle fluxes between sites $j$ and $j+1$,
\begin{equation}
  r_{j+1,j} \equiv 
  \frac{1}{2}
  \log 
  \left[
    \frac{
      \left( 1 + g {\nu}_J \right) {\nu}_j
    }{
      \left( 1 + g {\nu}_0 \right) {\nu}_{j+1}
    }
  \right] .
\label{eq:r_jp1_j_def}
\end{equation}
In terms of these the $\sigma$-dependence of $\hat{\Omega}$ may be
simplified to read 
\begin{widetext}
\begin{equation}
  \hat{\Omega} 
  \left( 
    \sigma , \nu
  \right) = 
  2 
  \sqrt{
    \left( 1 + g {\nu}_0 \right)
    \left( 1 + g {\nu}_J \right)
  } 
  \sum_{j=0}^{J-1}
  \left\{
    \sqrt{
      {\nu}_{j+1} {\nu}_{j}
    }
    \left[
      \cosh 
        \left( r_{j+1,j} \right) - 
      \cosh
      \left(
        {\sigma}_j - {\sigma}_{j+1} - r_{j+1,j}
      \right)
    \right]
  \right\} . 
\label{eq:Omega_hat_phys_genJ}
\end{equation}

The one-dimensional geometry we have assumed for the graph of
phosphorylation and dephosphorylation transitions makes possible the
definition of a function $\sigma \! \left( \nu \right)$ at each number
configuration, satisfying 
\begin{equation}
  {\sigma}_{j+1} \! \left( \nu \right) - 
  {\sigma}_j \! \left( \nu \right) = 
  - r_{j+1,j} , 
\label{eq:sigma_nu_def}
\end{equation}
(up to a convention for specifying $\bar{\sigma}$ at each $\nu$, which
we may take to be arbitrary).  $\sigma \! \left( \nu \right)$ is a
reference point for the {\em canonical} momentum coordinate $\sigma$,
and $\sigma - \sigma \!  \left( \nu \right)$ functions as the {\em
  kinematic momentum} for this system.

We may see this by introducing the Hamiltonian potential function 
\begin{equation}
  V \! 
  \left( 
    \nu
  \right) = 
  2 
  \sqrt{
    \left( 1 + g {\nu}_0 \right)
    \left( 1 + g {\nu}_J \right)
  } 
  \sum_{j=0}^{J-1}
  \sqrt{
    {\nu}_{j+1} {\nu}_{j}
  }
  \left[ 
    \cosh 
    \left( r_{j+1,j} \right) - 
    1 
  \right] , 
\label{eq:Omega_hat_pot_def}
\end{equation}
in terms of which 
\begin{equation}
  - \hat{\Omega} 
  \left( 
    \sigma , \nu
  \right) = 
  2 
  \sqrt{
    \left( 1 + g {\nu}_0 \right)
    \left( 1 + g {\nu}_J \right)
  } 
  \sum_{j=0}^{J-1}
  \sqrt{
    {\nu}_{j+1} {\nu}_{j}
  }
  \left\{
    \cosh
    \left[
      {\left( \sigma - \sigma \! \left( \nu \right) \right)}_j - 
      {\left( \sigma - \sigma \! \left( \nu \right) \right)}_{j+1}
    \right] - 
    1 
  \right\} - 
  V \! \left( \nu \right) . 
\label{eq:Omega_hat_phys_genJ_exp}
\end{equation}
\end{widetext}
To leading order the explicit $\cosh$ in
Eq.~(\ref{eq:Omega_hat_phys_genJ_exp}) is simply a quadratic form in
$\sigma - \sigma \! \left( \nu \right)$, with a matrix of inverse
masses defined by the remaining square-root terms.  When $\sigma -
\sigma \! \left( \nu \right) = 0$, Eq.~(\ref{eq:EOM_sigma_nu_form})
shows that ${\partial}_z \nu = 0$, verifying the interpretation of
$\sigma - \sigma \! \left( \nu \right)$ as the kinematic momentum.
Furthermore, if this kinematic momentum vanishes at any local minimum
of $V \! \left( \nu \right)$, Eq.~(\ref{eq:EOM_nu_sigma_form}) shows
that ${\partial}_z \sigma = 0$, hence ${\partial}_z \left( \sigma -
\sigma \! \left( \nu \right) \right) = 0$.  The local minima satisfy
$V \! \left( \nu \right) = 0$ and are attained only when each
$r_{j+1,j} = 0$ independently, because the $\cosh$ terms in
Eq.~(\ref{eq:Omega_hat_pot_def}) are never negative.  These are of
course exactly the (stable and saddle-point) mean-field solutions with
particle exchange between adjacent sites obeying detailed balance.  We
identify them in graphs below as the fixed points of the classical
diffusion equation.

\begin{figure}[ht]
  \begin{center} 
  \includegraphics[scale=0.5]{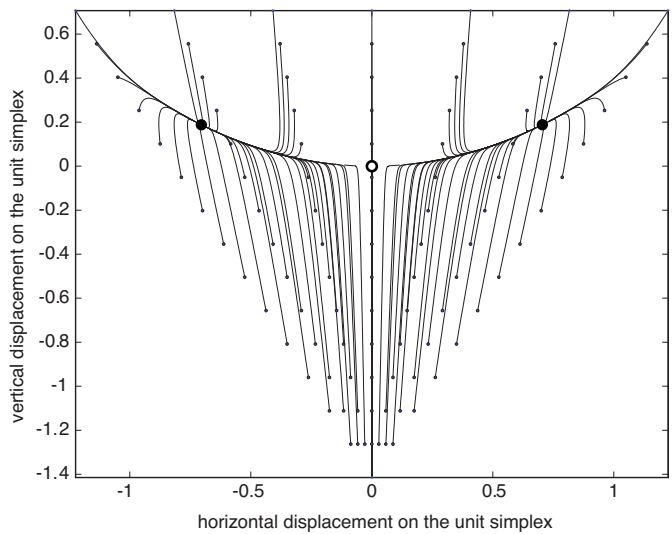}
  \caption{
  Classical flowfield of the particle numbers ${\nu}_j$ for $J=2$,
  shown on the simplex $\sum_{j=0}^2 {\nu}_j \equiv 1$.  Upper left
  corner is $n_0 = 1$, upper right corner is $n_2 = 1$, and bottom is
  $n_1 = 1$.  Dots are a selection of initial conditions, and
  classical diffusion solutions are the flowlines emanating from them.
  The uniform distribution ${\nu}_j = 1 / 3$ (open circle) is the
  saddle point, and the stable fixed points (heavy dots) are
  attractors.
    \label{fig:flowfields_psi_1} 
  }
  \end{center}
\end{figure}

It can be shown~\cite{Smith:evo_games:11} that, as long as the
quadratic expansion in $\sigma$ is a good approximation, and as long
as the effective mass terms implicit in
Eq.~(\ref{eq:Omega_hat_phys_genJ_exp}) are not a strong function of
$\nu$ (which we will verify), all stationary points of the action
closely approximate ordinary mechanical trajectories in the potential
$- V \! \left( \nu \right)$, with position coordinate $\nu$ and
kinematic momentum $\sigma - \sigma \! \left( \nu \right)$.  For the
classical solutions $\sigma \equiv 0$, shown in
Fig.~\ref{fig:flowfields_psi_1}, the unbounded trajectories are those
that originate in non-equilibrium initial conditions and converge
exponentially slowly on the saddle or stable fixed points.  The two
bounded trajectories, between the saddle and either stable fixed
point, travel along the saddle path of the potential, $- V \! \left(
\nu \right)$, which is bounded above by 0 and unbounded below, and
make up part of the {\em heteroclinic
  network}~\cite{Gluckheimer:het_cycles:88,Gluckheimer:dyn_sys:02} of
the associated hyperbolic system. 

For ordinary mechanical flow, we know that the full heteroclinic
network consists of trajectories running both ways between the stable
and saddle fixed points.  If this were a purely mechanical system, the
reverse trajectories would be strict time-reversal images of the
bounded classical diffusion trajectories.  (Note that an exact
reversal $\left( {\sigma}_j - {\sigma}_{j+1} \right) \rightarrow 2
r_{j+1,j} - \left( {\sigma}_j - {\sigma}_{j+1} \right)$ would also
leave $\hat{\Omega} = 0$.) Here, a small $\nu$-dependence of the
effective mass terms causes them to differ slightly from each other
and from the saddle path over $- V \! \left( \nu \right)$.

\begin{figure}[ht]
  \begin{center} 
  \includegraphics[scale=0.45]{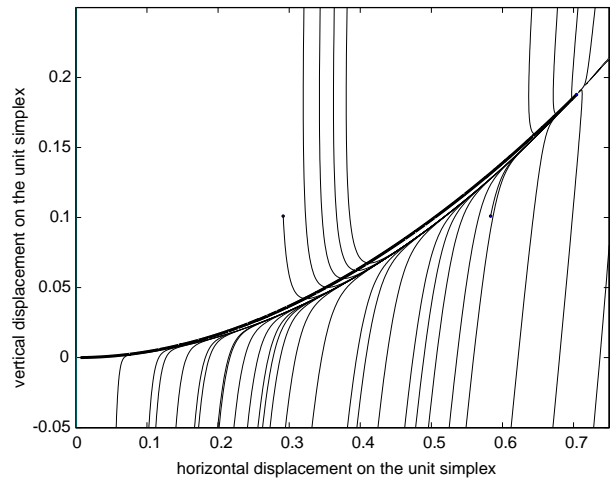}
  \caption{
  Flowfield of the instanton solution in $J=2$, for values $\xi = 1$,
  so $g \approx 4.0862$.  Fine lines are classical diffusion solution
  from Fig.~\ref{fig:flowfields_psi_1}, and bold line is the instanton.
    \label{fig:compare_flowfield_closeup_smooth} 
  }
  \end{center}
\end{figure}

In Fig.~\ref{fig:compare_flowfield_closeup_smooth}, we directly
compute the trajectory of the reverse bounded path, by integration
along the saddle instability of the equations of
motion~(\ref{eq:EOM_sigma_nu_form},\ref{eq:EOM_nu_sigma_form}).  The
fact that it nearly retraces the classical diffusive direction of
slowest flow checks the approximation that both trajectories are
dominated by the potential $- V \! \left( \nu \right)$ itself.

\begin{figure}[ht]
  \begin{center} 
  \includegraphics[scale=0.5]{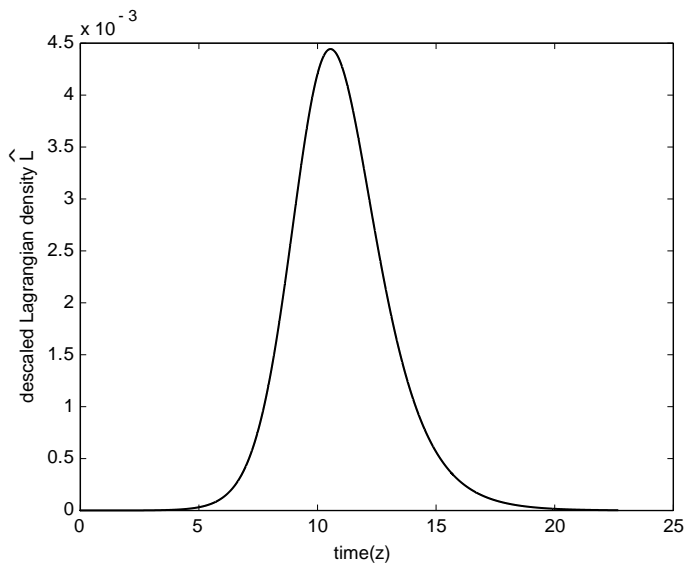}
  \caption{
  Linear-scale plot of the Lagrangian density per particle
  $\bar{\hat{L}}$ of the instanton trajectory.  A logarithmic plot of
  the same quantity (not shown) demonstrates exponential decay toward
  zero at early and late times.  
    \label{fig:instanton_L_linear} 
  }
  \end{center}
\end{figure}

For an instanton in a classical equilibrium field theory, the
conjugate and the kinematic momentum would be the same quantity.  Both
forward and reverse trajectories along the saddle path in the
potential $- V \! \left( \nu \right)$ would have locally minimum but
non-zero action, and in that sense both would be ``non-classical''
trajectories~\cite{Coleman:AoS:85}.  The distinctive feature of the
path integrals associated with master equations of the form we have
considered here is the offset $\sigma \! \left( \nu \right)$ from the
canonical momentum that appears in the kinetic term of the
Lagrangian~(\ref{eq:L_sigma_nu_form}) to the kinematic momentum.  This
offset is responsible for $S \equiv 0$ for all diffusion solutions,
including the bounded trajectories from saddle to stable fixed points,
and it approximately doubles the value of the
Lagrangian~(\ref{eq:L_sigma_nu_form}) along the reverse trajectories.
The value of this Lagrangian, along the numerically determined path of
Fig.~\ref{fig:compare_flowfield_closeup_smooth}, is shown in
Fig.~\ref{fig:instanton_L_linear}.  Like the Lagrangian for a
classical problem, it is positive definite, and approximates the WKB
integral for barrier escape, except with an extra factor of 2: $S
\approx 2 N \int \sqrt{d{\nu}^T m d\nu} \sqrt{2 V \! \left( \nu
  \right)}$, where $d\nu$ is a length element on the coordinate $\nu$,
and $m$ is the matrix of effective mass values implied by
Eq.~(\ref{eq:Omega_hat_phys_genJ_exp}).  This approximate form follows
simply from the nearly time-reverse character of escapes versus
classical paths of slowest-diffusion, and may be derived from the
original Gaussian-order approximation to such escapes by Onsager and
Machlup~\cite{Onsager:Machlup:53}.  Readers seeking a systematic
derivation, including the Onsager-Machlup small-fluctuation
approximation, may find these in Ref.~\cite{Smith:LDP_SEA:11} or
Ref.~\cite{Smith:evo_games:11}.  

\begin{figure}[ht]
  \begin{center} 
  \includegraphics[scale=0.45]{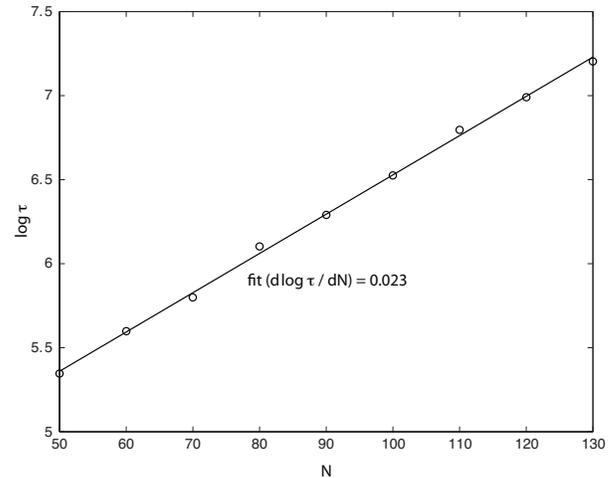}
  \caption{
  Plot comparing residence times to particle number for the parameters
  of Fig.~\ref{fig:instanton_L_linear}.  Slope of numerical estimates,
  which should correspond to $d S_0 / dN = \int dz \bar{\hat{L}}$ is
  best fit by $0.023$, while direct integration under the curve of
  Fig.~\ref{fig:instanton_L_linear} yields $0.0206$. 
    \label{fig:residence_vs_N} 
  }
  \end{center}
\end{figure}

For the parameter values of
Fig.~\ref{fig:compare_flowfield_closeup_smooth}, the integral $\int dz
\bar{\hat{L}}$ under the curve of Fig.~\ref{fig:instanton_L_linear}
converges to a value near $0.0206$.  Fig.~\ref{fig:residence_vs_N}
compares numerical estimates of the residence time in this model,
inverse to the rate $r_{\mbox{\scriptsize flip}}$ of
Eq.~(\ref{eq:flip_rate}), to particle number $N$.  The slope of the
logarithm of $1 / r_{\mbox{\scriptsize flip}}$ should be $d S_0 / dN =
\int dz \bar{\hat{L}}$, up to corrections decaying as $1/N$, and we
observe quantitative agreement with the numerical estimate of the
instanton to $\sim 10 \%$.

\begin{figure}[ht]
  \begin{center} 
  \includegraphics[scale=0.7]{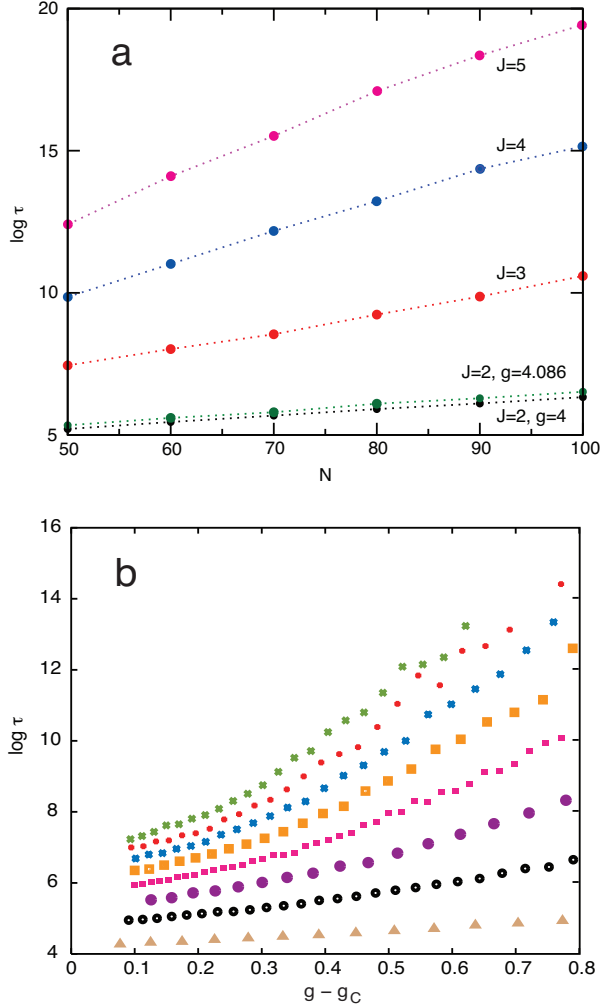}
  \caption{
  Plot of expected residence times $\tau$ from simulations in the
  symmetry-broken state as functions of $J$, $N$, and distance of the
  coupling strength from its critical value, $g - g_C$.  (a): $\log
  \tau$ vs. $N$ for $J \in \left\{ 2, \ldots , 5 \right\}$.  In
  simulations $I \equiv P$, $g = N / \sqrt{IP}$ is held constant, and
  $g - g_C$ is held equal to 1.  Green curve for $J = 2$ and $g =
  4.0862$ corresponds to the value $\xi \equiv 1$, which we have
  computed analytically, corresponding to the coefficient $d \log \tau
  / dN$ shown in Fig.~\ref{fig:residence_vs_N}.  (b): $\log \tau$
  vs. $g - g_C$ for values $J \in \left\{ 2, \ldots , 9 \right\}$
  (from bottom to top) at $N = 50$.
    \label{fig:figure7} 
  }
  \end{center}
\end{figure}

Although we do not pursue the analytic forms in this paper, the
dependence of the coefficient of $N$ (giving decay times\footnote{In
large-deviations terminology, this coefficient is called the
\emph{rate function}~\cite{Touchette:large_dev:09}.})  on number of
sites $J$ and the distance from criticality $g - g_C$ may also be
found from simulations.  Fig.~\ref{fig:figure7} symmarizes these
numerical results, showing that the dependence on small-integer $J$ is
roughly linear, and the dependence on $g - g_C < 1$ is weakly
nonlinear.

\subsection{Summary of semiclassical results}

We have seen that the classical ``action'' for this reaction-diffusion
theory, once constructed, yields quite nicely the two extremes of
behavior of interest in cooperative intermolecular phase transitions.
The classical stationary points coincide exactly with the usual
mass-action differential equations.  Corrections to these from cubic
and higher-order fluctuation effects are readily incorporated (for an
example, see Ref.~\cite{Smith:evo_games:11}), but to the resolution of
our simulations, we cannot identify a need for such corrections at
these parameter values, so we have not pursued them.

The nonclassical stationary trajectories are the projection from this
$\left( 2J+2 \right)$-dimensional configuration space, onto the
one-dimensional path most likely to destabilize the symmetry-broken
phase, which closely approximates the path of slowest diffusive
correction.  The methods shown here therefore provide a compact and
convenient way to estimate the escape trajectories and first-passage
times for even quite richly structured nonlinear diffusion processes
of this kind.  These methods, originally developed for applications to
reaction-diffusion theory, are increasingly finding applications in
epigenetics~\cite{Aurell:epigenetics:02,Roma:epigenetics:05} and
systems biology~\cite{Sinitsyn:CG_chem_nets:09} where particle numbers
may be small, making fluctuation effects important, while at the same
time the structure of the state space remains complex to describe.

The reduction to a one-dimensional system was assumed given, in the
treatment in Ref.~\cite{Bialek:memory:01} of a switching system
comparable to ours; we have shown here a systematic approach to
estimating such escape trajectories.  We have also verified that the
action of the instanton, easily numerically integrated once the
trajectory is found, produces both the correct $N$ scaling, and good
quantitative agreement, with the domain residence times.  We expect
that, while the treatment of the functional determinant will be more
difficult for metastable domains in first-order transitions, the
classical-level analysis comparable to ours will be similar, and
roughly as effective.

\section{Discussion and conclusions}
\label{sec:conclusions}

We have tried to idealize in a reasonable way a large class of
biomolecular signal transduction systems, and to apply the most
complete formalism available to decompose and quantitatively estimate
their properties as switches.  Our results thus combine a number of
technical advances in recognized domains, with several conceptual
insights relevant to the robustness and evolution of devices.  Of
necessity in a short treatment, our idealizations of feedback and
single timescales for all microscopic motion have abstracted away from
some important problems of connection with phenomenological models of
real regulatory protein systems.  

\subsection{Technical advances}

The operator treatments of gene-expression
switching~\cite{Sasai:gene_exp:03} extended traditional mass-action
models to include perturbative noise from first principles.  We have
further extended the operator methods to a path-integral treatment,
which adds an intuitive and computationally tractable approach to
large deviations.

We have demonstrated that expansion of weakly nonlinear stochastic
processes about Poisson backgrounds leads to very efficient
perturbative schemes for correcting the full probability distribution
(not very surprising in retrospect), and that in our particular
idealized model, the entire noise spectrum is driven by a single bare
Langevin field (perhaps somewhat more surprising, and not noticed
before).

Finally, we have shown that the nonlinear projection of the full
master equation onto the dominant trajectory participating in domain
flips approximately, but not exactly, reverses the unique trajectory
of slowest diffusive correction in the classical flow.  We have
recovered the exponential in $N$ characteristic of extensive
large-deviations scaling~\cite{Touchette:large_dev:09}, and shown how
to estimate the exact coefficient to refine the bounds of order unity
that are conventionally (and usually correctly) assumed in pure
scaling arguments~\cite{Bialek:memory:01}.

\subsection{Biological insights}

The most concrete of our results for biologists seeking to understand
the function of signal-transduction cascades and switches is that
particle number ($N$), as well as exogenous kinase ($I$) and
phosphatase ($P$) numbers, can be used to control the onset of
switching, and in cases of asymmetric topology, also the preferred
domain of the switch.  The control through $N$ offers a feedback from
gene expression into the function of the cascade, which apparently has
not been considered earlier.

We have demonstrated through the mode expansion in the neighborhood of
the phase transition, that only the lowest-eigenvalue collective
fluctuation of the diffusion operator induces the instability to
symmetry breaking, and scales the divergence in the noise spectrum.
For large $J$, we can also predict fluctuations in the symmetry-broken
phase because of a simplification induced by the fact that the system
senses only one boundary in the entire symmetry-broken phase. This
enables us to simplify an exact equation for the center-of-mass of the
system to predict fluctuations.

More abstractly, we have distinguished a mechanism for switching based
purely on population-level polarization of the protein pool, from
mechanisms which depend on limiting one or more transition rates
through restrictions on catalytic kinetics.  Polarization-based
mechanisms make the function of the switch dependent on its catalytic
topology and concentrations, but not on kinetic factors, a separation
that has been proposed as a route to modularity.  We hope that such
distinctions can at some point be incorporated in evolutionary models
that make quantitative use of the structure of phenotypically neutral
networks, where we hope they will explain at least part of the
ubiquity of multiple phosphorylation and nonspecific catalysis in
cascades of the type we have considered. 

\section{Acknowledgements} 

DES thanks Insight Venture Partners for support of this work. 
SK was supported by the swedish research council.


\begin{thebibliography}{10}

\bibitem{vonNeumann:problog:56}
J.~von Neumann.
\newblock Probabilistic logics and the synthesis of reliable organisms from
  unreliable components.
\newblock {\em Automata Studies}, 34:43--, 1956.

\bibitem{vonDassow:SPN:00}
G.~von Dassow, E.~Meir, E.~M. Munro, and G.~M. Odell.
\newblock The segment polarity network is a robust developmental module.
\newblock {\em Nature}, 406:188, 2000.

\bibitem{Sauro:proteomics:04}
H.~M. Sauro.
\newblock The computational versatility of proteomic signaling networks.
\newblock {\em Current Proteomics}, 1:67, 2004.

\bibitem{Ferrell:xenopus:99}
J.~E.~Jr. Ferrell.
\newblock Xenopus oocyte maturation: new lessons from a good egg.
\newblock {\em Bioessays}, 21:833, 1999.

\bibitem{Ferrell:switch:99}
J.~E.~Jr. Ferrell.
\newblock Building a cellular switch: more lessons from a good egg.
\newblock {\em Bioessays}, 21:866, 1999.

\bibitem{Tyson:CellCycle:02}
J.~J. Tyson, A.~Csikasz-Nagy, and B.~Novak.
\newblock The dynamics of cell cycle regulation.
\newblock {\em Bioessays}, 24:1095, 2002.

\bibitem{Sasai:gene_exp:03}
M.~Sasai and P.~Wolynes.
\newblock Stochastic gene expression as a many-body problem.
\newblock {\em Proc.~Nat.~Acad.~Sci.~USA}, 100:2374--2379, 2003.

\bibitem{Goldbeter:hypersensitivity:81}
A.~Goldbeter and D.~E. Koshland.
\newblock An amplified sensitivity arising from covalent modification in
  biological systems.
\newblock {\em Proc.~Nat.~Acad.~Sci.~USA}, 78:6840--6844, 1981.

\bibitem{Huang:MAPK:96}
C.~Y. Huang and J.~E.~Jr. Ferrell.
\newblock Ultrasensitivity in the mitogen-activated protein kinase cascade.
\newblock {\em Proc.~Nat.~Acad.~Sci.~USA}, 93:10078, 1996.

\bibitem{Tyson:Sniffers:03}
J.~J. Tyson, K.~C. Chen, and B.~Novak.
\newblock Sniffers, buzzers, toggles and blinkers: dynamics of regulatory and
  signaling pathways in the cell.
\newblock {\em Curr.~Opin.~Cell Biol.}, 15:221, 2003.

\bibitem{Ferrell:bistability:01}
J.~E. Ferrell and W.~Xiong.
\newblock Bistability in cell signaling: How to make continuous processes
  discontinuous, and reversible processes irreversible.
\newblock {\em Chaos}, 11:227, 2001.

\bibitem{Lisman:bistability:85}
J.~E. Lisman.
\newblock A mechanism for memory storage insensitive to molecular turnover: a
  bistable autophosphorylating kinase.
\newblock {\em PNAS}, 82:3055, 1985.

\bibitem{Bialek:memory:01}
W.~Bialek.
\newblock Stability and noise in biochemical switches.
\newblock In T.~K. Leen, T.~G. Dietterich, and V.~Tresp, editors, {\em Advances
  in Neural Information Processing}, Cambridge, 2001. MIT Press.

\bibitem{Kultz:stress_signals:98}
D.~\protect{K\"{u}ltz} and M.~Burg.
\newblock Evolution of osmotic stress signaling via map kinase cascades.
\newblock {\em J.~Exp.~Biol.}, 201:3015, 1998.

\bibitem{Gunawardena:switch:05}
Jeremy Gunawardena.
\newblock Multisite protein phosphorylation makes a good threshold but can be a
  poor switch.
\newblock {\em Proc.~Nat.~Acad.~Sci.~USA}, 102:14617--14622, 2005.

\bibitem{Salazar:phosphate:07}
{Salazar, C. and H\"{o}fer, T.}
\newblock Versatile reglation of multisite protein phosphorylation by the order
  of phosphate processing and protein-protein interactions.
\newblock {\em FEBS journal}, 274:1046--1061, 2007.

\bibitem{Kreegipuu:phosphobase:99}
A.~Kreegipuu, N.~Blom, and S.~Brunak.
\newblock Phosphobase, a database of phosphorylation sites: release 2.0.
\newblock {\em Nucleic Acids Res.}, 27:237, 1999.

\bibitem{Markevich:signalling:04}
N.~I. Markevich, J.~B. Hoek, and B.~N. Kholodenko.
\newblock Signaling switches and bistability arising from multisite
  phosphorylation in protein kinase cascades.
\newblock {\em J. ~Cell ~Biol.}, 164:353--359, 2004.

\bibitem{Craciun:bistability:06}
G.~Craciun, Y.~Tang, and M.~Feinberg.
\newblock Understanding bistability in complex enzyme-driven reaction networks.
\newblock {\em Proc.~Nat.~Acad.~Sci.~USA}, 103:8697--8702, 2006.

\bibitem{Ancel:mod_RNA:00}
L.~W. Ancel and W.~Fontana.
\newblock Plasticity, evolvability and modularity in rna.
\newblock {\em J.~Exp.~Zool.~(Mol.~Dev.~Evol.)}, 288:242--283, 2000.

\bibitem{Fontana:evo_devo_RNA:02}
W.~Fontana.
\newblock Modeling `evo-devo' with rna.
\newblock {\em Bioessays}, 24:1164--1177, 2002.

\bibitem{Takahashi:spatio-temporal:10}
K.~Takahasi, S.~Tanase-Nicola, and P.~Rein~ten Wolde.
\newblock Spatio-temporal correlations can drastically change the response of a
  mapk pathway.
\newblock {\em PNAS}, 107:2473--2478, 2010.

\bibitem{Kholodenko:negative:00}
B.~N. Kholodenko.
\newblock Negative feedback and ultrasensitivity can bring about oscillations
  in the mitogen-activated protein kinase cascades.
\newblock {\em Eur. ~J. ~Biochem}, 267:1583--1588, 2000.

\bibitem{Heinrich:mathmodels:02}
R.~Heinrich, B.~G. Neel, and T.~A. Rapoport.
\newblock Mathematical models of protein kinase signal transduction.
\newblock {\em Mol.~Cell.}, 9:957--970, 2002.

\bibitem{Wang:stochastic:06}
X.~Wang, N.~Hao, H.~G. Dohlman, and T.~C. Elston.
\newblock Bistability, stochasticity and oscillations in the mitogen-activated
  protein kinase cascade.
\newblock {\em Biophys. ~J.}, 90:1961--1978, 2006.

\bibitem{Kapuy:mol_switches:09}
Orsolya Kapuy, Debashis Barik, Maria Rosa~Domingo Sananes, John~J. Tyson, and
  B\'{e}la Nov\'{a}k.
\newblock Bistability by multiple phosphorylation of regulatory proteins.
\newblock {\em Prog.~Biophys.~Mol.~Biol.}, 100:47--56, 2009.

\bibitem{Novak:CellCycle:01}
B.~Novak, Z.~Pataki, A.~Ciliberto, and J.~J. Tyson.
\newblock Mathematical model of the cell division cycle of fission yeast.
\newblock {\em Chaos}, 11:277, 2001.

\bibitem{Paulsson:noise:01}
J.~Paulsson and M.~Ehrenberg.
\newblock Noise in a minimal regulatory network: plasmid copy number control.
\newblock {\em Q.~Rev.~Biophys.}, 34:1, 2001.

\bibitem{Krishnamurthy:Signaling:07}
Supriya Krishnamurthy, Eric Smith, David~C. Krakauer, and Walter Fontana.
\newblock The stochastic behavior of a molecular switching circuit with
  feedback.
\newblock {\em Biology Direct}, 2:13, 2007.
\newblock doi:10.1186/1745-6150-2-13.

\bibitem{Ferrell:MAPK:97}
J.~E. Ferrell and R.~R. Bhatt.
\newblock Mechanistic studies of the dual phosphorylation of mitogen-activated
  protein kinase.
\newblock {\em J. Biol. Chem.}, 272:19008--19016, 1997.

\bibitem{Burack:nonprocessive:97}
W.~R. Burack and T.~W. Sturgill.
\newblock The activating dual-phosphorylation of mapk by mek is nonprocessive.
\newblock {\em Biochemistry}, 36:5929--5933, 1997.

\bibitem{Gaudet:compendium:05}
S.~Gaudet, K.~A. Janes, J.~Albeck, E.~A. Pace, D.~A. Lauffenburger, and P.~K.
  Sorger.
\newblock A compendium of signals and responses triggered by prodeath and
  prosurvival cytokines.
\newblock {\em Mol.~Cell Proteomics}, 10:1569--1590, 2005.

\bibitem{Kravanja:hprK:99}
M.~Kravanja, R.~Engelmann, V.~Dossonnet, M.~Bl\"{u}ggel, H.~E. Meyer, R.~Frank,
  A.~Galinier, J.~Deutscher, N.~Schnell, and W.~Hengstenberg.
\newblock The hprk gene of enterococcus faecalis encodes a novel bifunctional
  enzyme: the hpr kinase/phosphatase.
\newblock {\em Mol.~Microbiol.}, 31:59--66, 1999.

\bibitem{Ninfa:protein:91}
A.~J. Ninfa.
\newblock Protein phosphorylation and the regulation of cellular processes by
  the homologous two-component regulatory systems of bacteria.
\newblock {\em Genet.~Eng.}, 13:39--72, 1991.

\bibitem{Coleman:AoS:85}
Sidney Coleman.
\newblock {\em Aspects of symmetry}.
\newblock Cambridge, New York, 1985.

\bibitem{Mattis:RDQFT:98}
Daniel~C. Mattis and M.~Lawrence Glasser.
\newblock The uses of quantum field theory in diffusion-limited reactions.
\newblock {\em Rev.~Mod.~Phys}, 70:979--1001, 1998.

\bibitem{Eyink:action:96}
Gregory~L. Eyink.
\newblock Action principle in nonequilibrium statistical dynamics.
\newblock {\em Phys.~Rev.~E}, 54:3419--3435, 1996.

\bibitem{Cardy:FTNEqSM:99}
J.~Cardy.
\newblock Field theory and non-equilibrium statistical mechanics.
\newblock 1999.
\newblock http://www-thphys.physics.ox.ac.uk/users/JohnCardy/home.html.

\bibitem{Smith:LDP_SEA:11}
Eric Smith.
\newblock Large-deviation principles, stochastic effective actions, path
  entropies, and the structure and meaning of thermodynamic descriptions.
\newblock {\em Rep.~Prog.~Phys.}, 74:046601, 2011.
\newblock http://arxiv.org/submit/199903.

\bibitem{Sinitsyn:CG_chem_nets:09}
N.~A. Sinitsyn, Nicalas Hengartner, and Ilya Nemenman.
\newblock Adiabatic coarse-graining and simulations of stochastic biochemical
  networks.
\newblock {\em Proc.~Nat.~Acad.~Sci.~USA}, 106:10546--10551, 2009.

\bibitem{Gillespie:QTMP:94}
Daniel~T. Gillespie.
\newblock Why quantum mechanics cannot be formulated as a markov process.
\newblock {\em Phys.~Rev.~A}, 49:1607--1612, 1994.

\bibitem{Aurell:epigenetics:02}
E.~Aurell and K.~Sneppen.
\newblock Epigenetics as a first exit problem.
\newblock {\em Phys.~Rev.~Lett.}, 88:048101:1--4, 2002.

\bibitem{Touchette:large_dev:09}
Hugo Touchette.
\newblock The large deviation approach to statistical mechanics.
\newblock {\em Phys.~Rep.}, 478:1--69, 2009.
\newblock arxiv:0804.0327.

\bibitem{Roma:epigenetics:05}
David~Marin Roma, Ruadhan~A. O'Flanagan, Andrei~E. Ruckenstein, and Anirvan~M.
  Sengupta.
\newblock Optimal path to epigenetic switching.
\newblock {\em Phys.~Rev.~E}, 71:011902:1--5, 2005.

\bibitem{Smith:evo_games:11}
Eric Smith and Supriya Krishnamurthy.
\newblock Symmetry and collective fluctuations in evolutionary games.
\newblock 2011.
\newblock SFI working paper \#11-03-010.

\bibitem{Weinberg:QTF_II:96}
Steven Weinberg.
\newblock {\em {\protect The quantum theory of fields, Vol.~II: Modern
  applications}}.
\newblock Cambridge, New York, 1996.

\bibitem{Cardy:Instantons:78}
J.~L. Cardy.
\newblock Electron localisation in disordered systems and classical solutions
  in ginzburg-landau field theory.
\newblock {\em J.~Phys.~C}, 11:L321 -- L328, 1987.

\bibitem{Onsager:Machlup:53}
L.~Onsager and S.~Machlup.
\newblock Fluctuations and irreversible processes.
\newblock {\em Phys.~Rev.}, 91:1505, 1953.

\bibitem{Gluckheimer:het_cycles:88}
John Gluckheimer and Philip Holmes.
\newblock Structurally stable heteroclinic cycles.
\newblock {\em Math.~Proc.~Cam.~Phil.~Soc.}, 103:189--192, 1988.

\bibitem{Gluckheimer:dyn_sys:02}
John Gluckheimer and Philip Holmes.
\newblock {\em Nonlinear Oscillations, Dynamical Systems, and Bifurcations of
  Vector Fields (Applied Mathematical Sciences vol.42)}.
\newblock Springer, Berlin, 2002.

\bibitem{Goldstein:ClassMech:01}
Herbert Goldstein, Charles~P. Poole, and John~L. Safko.
\newblock {\em Classical Mechanics}.
\newblock Addison Wesley, New York, third edition, 2001.

\bibitem{Kamenev:DP:01}
Alex Kamenev.
\newblock Keldysh and doi-peliti techniques for out-of-equilibrium systems.
\newblock pages arXiv:cond--mat/0109316v2, 2001.
\newblock Lecture notes presented at Windsor NATO school on ``Field Theory of
  Strongly Correlated Fermions and Bosons in Low-Dimensional Disordered
  Systems'' (August 2001).

\bibitem{Keldysh::65}
L.~V. Keldysh.
\newblock Diagram technique for nonequilibrium processes.
\newblock {\em Sov.~Phys.~JETP}, 20:1018, 1965.

\bibitem{Martin:MSR:73}
P.~C. Martin, E.~D. Siggia, and H.~A. Rose.
\newblock Statistical dynamics of classical systems.
\newblock {\em Phys.~Rev.~A}, 8:423--437, 1973.

\end{thebibliography}

\end{document}